\documentclass[11pt,a4paper]{article}

\usepackage[utf8]{inputenc}
\usepackage[T1]{fontenc}
\usepackage{lmodern}
\usepackage{microtype}
\usepackage[margin=1in]{geometry}
\usepackage{amsmath,amssymb,amsthm,bm}
\usepackage{graphicx}
\usepackage{booktabs}
\usepackage{array}
\usepackage{tabularx}
\usepackage{ragged2e}
\usepackage{algorithm}
\usepackage{algpseudocode}
\usepackage{xcolor}
\usepackage{authblk}
\usepackage{float}
\usepackage[section]{placeins}
\usepackage[numbers,sort&compress]{natbib}
\usepackage{hyperref}
\hypersetup{
  colorlinks=true,
  urlcolor=blue,
  linkcolor=blue,
  citecolor=blue,
  pdftitle={When a positive SIMP density floor is not enough: solver admissibility and guarded floor selection in matrix-free 3D topology optimization},
  pdfauthor={Shaoliang Yang, Jun Wang, Yunsheng Wang},
  pdfkeywords={topology optimization, SIMP, density floor, solver admissibility, matrix-free finite elements, geometric multigrid, residual acceptance}
}
\newcolumntype{L}[1]{>{\RaggedRight\arraybackslash}p{#1}}
\newcolumntype{Y}{>{\RaggedRight\arraybackslash}X}

\newtheorem{proposition}{Proposition}
\theoremstyle{remark}
\newtheorem{remark}{Remark}

\newcommand{\bK}{\bm{K}}
\newcommand{\bu}{\bm{u}}
\newcommand{\bff}{\bm{f}}
\newcommand{\rhoMin}{\rho_{\min}}
\newcommand{\rr}{\rho}

\title{\textbf{When a positive SIMP density floor is not enough:\\ solver admissibility and guarded floor selection in\\ matrix-free 3D topology optimization}}

\author[1]{Shaoliang Yang}
\author[1]{Jun Wang\thanks{Corresponding author. E-mail: jwang22@scu.edu}}
\author[1]{Yunsheng Wang}
\affil[1]{Department of Mechanical Engineering, Santa Clara University, Santa Clara, CA 95053, USA}
\date{}

\begin{document}
\maketitle

\begin{abstract}
In a matrix-free geometric-multigrid FGMRES solver for three-dimensional SIMP topology optimization, a reported converged solve is not always a converged solve. On four of 102 held-out states, the projected residual used for stopping falls below \(10^{-6}\) while a recomputed true residual is \(1.35\)--\(49.5\) times the tolerance; in an unguarded optimization trajectory, 22 of 40 state solves reach the iteration cap without raising an error. We formulate floor selection as a verified control problem: probe the frozen state at the original floor, use two residual features to choose the first attempted floor, and accept no solution until \(\|\bff-\bK\bu\|/\|\bff\|\le10^{-6}\) is recomputed. The two-feature rule matches 98 of 102 reference classifications; the residual guard detects the four missed escalations, and all 102 selected solves satisfy the tolerance. Relative to always using a \(10^{-3}\) floor, the policy preserves the original operator on 24 admissible states and avoids mean compliance and gradient changes of \(31.0\%\) and \(0.340\) on those severe random states, and \(0.48\%\) and \(0.008\) on seven optimized designs, at \(2.5\) times the mean wall time. In a 12-state subset of the held-out states, eight still require escalation at the conventional floor \(10^{-6}\). In a nine-state control with the preconditioner's adaptive components disabled, every failure is visible and no false acceptance occurs, tying the stopping-estimate drift to the iterate-dependent preconditioner. The recomputed residual is the correctness safeguard; the probe and floor ladder govern an implementation-specific cost--fidelity tradeoff.
\end{abstract}


\section{Introduction}

Density-based topology optimization is mature enough that the limiting question is often not how to formulate a compliance problem, but whether the finite-element state solves remain trustworthy as designs approach large, three-dimensional, high-contrast regimes. The SIMP lineage supplies the material interpolation framework \citep{bendsoe1988homogenization,bendsoe1999material,bendsoe2004book}, while filters, projection methods, compact benchmark codes, and recent analysis make density methods reproducible and mathematically better understood \citep{bourdin2001filters,lazarov2011filters,guest2004minimum,andreassen201188,ferrari2020top99neo,sigmund2013review,papadopoulos2025simp}. At practical 3D scale the repeated elasticity solve remains the bottleneck, and distributed-memory frameworks, PETSc implementations, Krylov recycling, multigrid preconditioners, GPU codes, and matrix-free operators have all been developed to make those solves feasible \citep{wang2007krylov,aage2013parallel,amir2014multigrid,aage2015petsc,aage2017giga,wu2016system,traff2023gpu,zhao2024gpu,qi2025gpu,wang2025matrixfree}.

Almost all of that work assumes the state solve either succeeds or fails visibly. This paper starts from a case where it does neither. In ersatz-material SIMP, void regions are kept weakly stiff by a positive density floor \(\rhoMin\). At the modeling level a positive floor is usually treated as sufficient to avoid a singular stiffness operator; it does not establish that a particular matrix-free multigrid hierarchy will reach the requested residual tolerance. We call a floor \emph{solver-admissible} for a given state, hierarchy, tolerance, and iteration budget when the solve actually attains that tolerance. A floor can be mathematically positive and solver-inadmissible.

The failure that motivates this paper is worse than a stalled solve, because a stalled solve is visible. On four of 102 held-out states, the right-preconditioned FGMRES iteration reported convergence while the recomputed unpreconditioned residual was \(1.35\times\) to \(49.5\times\) the requested tolerance (Figure~\ref{fig:policy-guard}b). The projected residual estimate that the Krylov method minimizes and the residual that certifies the equilibrium solution had separated. Inside an optimization loop the same separation is invisible: in a 40-iteration cantilever run at the original floor, the state solve saturates the 300-iteration budget from outer iteration 19 onward, and the recorded compliance oscillates between 0.484 and 4.502 while the optimizer continues (Figure~\ref{fig:in-loop}). Compliance and sensitivities computed from such states are not equilibrium quantities, but nothing in the workflow says so.

The reported experiments use a stringent original floor, \(\rho_0=10^{-12}\), because it makes inadmissibility easy to reach and therefore easy to study. The phenomenon is not confined to that choice: at \(\rho_0=10^{-6}\), eight of 12 states in a post-hoc sensitivity subset still require escalation before a solve is accepted (Section~\ref{sec:robust}). The problems, meshes, loads, and frozen states behind these numbers are specified in Section~\ref{sec:formulation} and drawn in Figure~\ref{fig:setup}.

Neither obvious workaround is satisfactory. Leaving the original floor unchanged lets hard low-density states fail, sometimes silently. Raising the floor globally improves convergence but changes the operator whose compliance and sensitivities are being evaluated: on the 24 held-out states where the original floor was admissible, substituting a fixed \(10^{-3}\) floor changes compliance by \(31.0\%\) on average and the solid-element compliance gradient by a mean relative \(\ell_2\) norm of \(0.340\) (\(55.2\%\) and \(0.555\) at \(10^{-2}\)), although on optimized designs the same substitution is far milder (Section~\ref{sec:perturb}). The useful question is therefore not whether a larger floor helps, but whether the original floor is admissible for the current frozen state, and, if not, what the smallest tested intervention is that restores a verified solve.

Prior work supplies the surrounding pieces. Large-scale 3D topology optimization establishes the need for scalable state solvers \citep{aage2013parallel,aage2015petsc,aage2017giga,yago2022comparative}; multigrid topology-optimization work shows that hierarchy details govern robustness \citep{amir2014multigrid,herrero2023adaptive,kruger2025orthotropic}; GPU and matrix-free studies reduce memory pressure and expose stack-specific failure modes \citep{wu2016system,traff2023gpu,padhi2023gpu,zhao2024gpu,qi2025gpu,wang2025matrixfree,yang2026fused,yang2026gmg}; mixed-precision solver theory explains why conditioning and residual evaluation matter when lower precision enters a hierarchy \citep{carson2017new,haidar2020tensor,higham2022mixed,mccormick2021mixedmg,tsai2023three}; and low-density pathology studies show that weak void stiffness creates numerical artifacts \citep{yoon2005ecp,dalklint2021buckling,zhang2021nonersatz}. To our knowledge, none of them provides an acceptance-guarded procedure for deciding when to preserve or escalate the SIMP floor, or reports the projected-versus-true residual acceptance failure in a topology-optimization setting.

The contributions are:
\begin{enumerate}\setlength{\itemsep}{1pt}
  \item evidence that a projected-residual stopping test can disagree with a recomputed equilibrium residual on SIMP states, both for frozen solves and inside an optimization loop;
  \item a guarded floor-selection policy that preserves the original floor whenever it is admissible and escalates only under residual evidence, with acceptance always gated on a recomputed true residual;
  \item a held-out evaluation of that policy against 300-iteration reference classifications and against fixed-floor baselines, including the cost the preservation-first choice actually incurs and the operator perturbation it avoids.
\end{enumerate}

\paragraph{Relation to prior implementations.}
The matrix-free elasticity operator and geometric-multigrid hierarchy used as numerical infrastructure are described in prior implementation preprints \citep{yang2026fused,yang2026gmg}. Those studies evaluate operator throughput, hierarchy variants, reduced precision, and visibly nonconvergent solves. The present work asks a different numerical question: whether a solve that satisfies the implementation's projected-residual stopping test also satisfies a freshly evaluated equilibrium residual, and how the density floor should be selected when it does not. No new kernel or multigrid hierarchy is claimed here. The optimized density fields of Section~\ref{sec:transfer} are used only as fixed transfer inputs, with source-solve limitations reported explicitly. The residual semantics, floor ladder, acceptance criterion, and solver variants needed to evaluate the present claims are specified in Section~\ref{sec:method}.

\paragraph{Organization.}
Section~\ref{sec:related} positions the work. Section~\ref{sec:method} defines solver admissibility, the probe features, and the guarded policy. Section~\ref{sec:results} reports the evidence, beginning with the acceptance failure itself. Section~\ref{sec:discussion} discusses when the policy is worth its cost, what mechanism the ablations support, and what the evidence does not cover. Appendix~\ref{app:repro} documents reproduction; Appendix~\ref{app:supp} holds supplementary tables and figures.

\section{Related Work}\label{sec:related}

\paragraph{Density methods and the ersatz floor.}
Density-based topology optimization descends from the homogenization approach of \citet{bendsoe1988homogenization} and the material-interpolation (SIMP) schemes analysed by \citet{bendsoe1999material}, consolidated in the monograph of \citet{bendsoe2004book}. The regularization practice this paper inherits is settled: \citet{bourdin2001filters} established density filtering, \citet{lazarov2011filters} recast the filter as a Helmholtz-type PDE, and \citet{guest2004minimum} introduced nodal projection for minimum length scale; compact reference implementations \citep{andreassen201188,ferrari2020top99neo} made the pipeline reproducible; level-set formulations \citep{wang2003levelset,allaire2004levelset} and the surveys of \citet{sigmund2013review} and \citet{yago2022comparative} chart the alternatives; and \citet{papadopoulos2025simp} has recently sharpened the mathematical status of SIMP compliance minimization. The weak-material floor itself is known to be more than a formality on the modeling side: \citet{yoon2005ecp} replaced low-density elements with an element-connectivity parameterization precisely because ersatz elements misbehave under geometric nonlinearity, \citet{dalklint2021buckling} showed that weak ersatz material can generate artificial buckling modes in stability-constrained design, and \citet{zhang2021nonersatz} constructed a non-ersatz formulation to avoid the weak phase altogether. These works treat the floor as a modeling liability and change the formulation to escape it. We change nothing in the formulation and ask the complementary, solver-facing question: when is a given positive floor operationally admissible for the hierarchy actually being used?

\paragraph{Scalable solvers for the repeated state problem.}
The cost of the repeated equilibrium solve has its own lineage. \citet{wang2007krylov} treated the state solves of a design loop as a sequence of related linear systems, recycling Krylov subspaces across design iterations and controlling how accurately each system needs to be solved; \citet{amir2014multigrid} showed that multigrid-preconditioned CG makes the 3D state solve affordable, with performance governed by the interpolated coefficient field rather than by definiteness alone --- a dependence that adaptive and orthotropic multigrid studies later examined in detail \citep{herrero2023adaptive,kruger2025orthotropic}. Distributed-memory practice was established by the parallel framework of \citet{aage2013parallel} and the open PETSc implementation of \citet{aage2015petsc}, culminating in the giga-voxel morphogenesis of \citet{aage2017giga}. On single machines, \citet{wu2016system} demonstrated high-resolution topology optimization with a GPU multigrid solver, \citet{traff2023gpu} published compact GPU-accelerated codes and applications, \citet{zhao2024gpu} and \citet{qi2025gpu} extended GPU solvers to fiber-reinforced composites, \citet{padhi2023gpu} combined GPU computing with homogenization-based multigrid, and \citet{wang2025matrixfree} released a matrix-free implementation for large-scale 3D problems. Throughout this line, a solve whose convergence flag is set is treated as solved; the present paper is about the cases in which that assumption fails.

\paragraph{Matrix-free operators on accelerators.}
Matrix-free finite-element computing supplies the operator technology. The MFEM library \citep{anderson2021mfem} organized matrix-free high-order operators for GPUs; \citet{davydov2020matrixfree} applied matrix-free geometric multigrid to finite-strain problems; low-order-refined preconditioning made high-order operators preconditionable end-to-end on GPUs \citep{franco2020lor,pazner2023lor}, with production-scale hydrodynamics experience reported by \citet{vargas2022matrixfree}; \citet{sun2020vectorization} analysed the vectorization of matrix-free kernels; and \citet{bohm2025galerkin} studied large-scale multigrid with adaptive Galerkin coarsening. The shared lesson is that on accelerators the operator representation and the preconditioner must be designed together. That is the setting here; the hierarchy itself is held fixed, and its acceptance semantics, not its construction, are the subject.

\paragraph{Krylov stopping estimates and finite-precision residuals.}
\citet{saad1993fgmres} introduced the flexible GMRES iteration used here, in which the quantity that stops the solver is the projected least-squares residual of the Krylov subspace; the standard texts \citep{saad2003iterative,briggs2000multigrid,trottenberg2000multigrid} and multigrid libraries \citep{naumov2015amgx,bell2023pyamg} define the surrounding practice. That an inexpensive residual estimate can drift from the explicitly computed residual in finite precision is classical, and modern mixed-precision analysis makes the point quantitative: \citet{carson2017new} and \citet{carson2018three} gave conditioning-dependent bounds on the accuracy that iterative refinement can attain across two and three precisions, \citet{mccormick2021mixedmg} carried the rounding-error analysis to multigrid, \citet{tsai2023three} and \citet{haidar2020tensor} brought the machinery to GPU solvers, and the surveys of \citet{higham2022mixed}, \citet{abdelfattah2021survey}, and \citet{kashi2026mixed} organize the field. The conditioning-times-roundoff quantity that appears in these analyses is exactly the \(\kappa\varepsilon\) indicator measured in Section~\ref{sec:conditioning}. The contribution tested below is therefore not a new GPU kernel or hierarchy. It is evidence, within the stated 3D SIMP setting and tolerance, that this known residual gap can reach a stopping decision and propagate into an optimization loop, together with a floor-selection procedure whose final decision is verified by a recomputed residual.

\section{Method}\label{sec:method}

We formulate floor selection as a verified intervention around a fixed matrix-free GMG-FGMRES equilibrium solve: preserve the original floor when an independently recomputed residual accepts it, otherwise choose the first tested ladder floor that produces an accepted solve. The material model and multigrid hierarchy are held fixed.

\subsection{Problem formulation, discretization, and benchmarks}\label{sec:formulation}

The design problem is standard minimum-compliance topology optimization for linear isotropic elasticity in three dimensions. On a domain \(\Omega\) with Dirichlet boundary \(\Gamma_D\) and traction boundary \(\Gamma_N\), the displacement field satisfies
\begin{equation}
  -\nabla\!\cdot\!\bm{\sigma}(\bu)=\bm{0}\ \text{in }\Omega,\qquad
  \bm{\sigma}=\mathbb{C}(E)\!:\!\bm{\varepsilon}(\bu),\qquad
  \bu=\bm{0}\ \text{on }\Gamma_D,\qquad
  \bm{\sigma}\bm{n}=\bm{t}\ \text{on }\Gamma_N,
  \label{eq:elasticity}
\end{equation}
with \(\mathbb{C}(E)\) the isotropic elasticity tensor for Young's modulus \(E\) and Poisson's ratio \(\nu\). After discretization the optimization problem is
\begin{equation}
  \min_{\rr\in[0,1]^{n_e}}\ C=\bff^{\top}\bu
  \quad\text{subject to}\quad
  \bK(\bar{\rr})\bu=\bff,\qquad
  \textstyle\sum_e \bar{\rr}_e \le V^{*} n_e ,
  \label{eq:to_problem}
\end{equation}
where \(\bar{\rr}\) is the physical density obtained from the design variable \(\rr\) by a linear-hat density filter of radius \(r_{\min}\) followed by the smoothed Heaviside projection
\begin{equation}
  \bar{\rr}_e=\frac{\tanh(\beta\eta)+\tanh\bigl(\beta(\tilde{\rr}_e-\eta)\bigr)}
                   {\tanh(\beta\eta)+\tanh\bigl(\beta(1-\eta)\bigr)},\qquad \eta=0.5 ,
  \label{eq:projection}
\end{equation}
and the element modulus follows the ersatz SIMP law of Eq.~\eqref{eq:floor_scaling} below. Compliance sensitivities are the usual self-adjoint expression \(\partial C/\partial\bar{\rr}_e=-p_{\mathrm{SIMP}}(1-\rhoMin)\bar{\rr}_e^{\,p_{\mathrm{SIMP}}-1}\bu_e^{\top}\bK_e\bu_e\), chained back through Eq.~\eqref{eq:projection} and the filter. The design update is the optimality-criteria rule with a move limit and a bisection on the volume multiplier \citep{bendsoe2004book,andreassen201188}. None of this is modified anywhere in the paper; the object of study is the equilibrium solve in Eq.~\eqref{eq:to_problem}.

\paragraph{Discretization and units.}
All meshes are structured grids of trilinear eight-node hexahedra with geometrically cubic elements. The element stiffness is formed once on the reference cube with \(E_0=1\) and \(\nu=0.3\) using \(2\times2\times2\) Gauss integration and is reused for every element, so densities enter only through the scalar factor \(E_e\) of Eq.~\eqref{eq:floor_scaling}. Loads are normalized to unit magnitude: point loads carry \(\|\bm{f}\|=1\) and distributed loads a unit pressure converted to consistent nodal forces by tributary area. Compliance values are therefore dimensionless quantities of this normalized system, comparable across the runs reported here but not to a dimensional design.

\begin{figure}[!tbp]
  \centering
  \includegraphics[width=\linewidth]{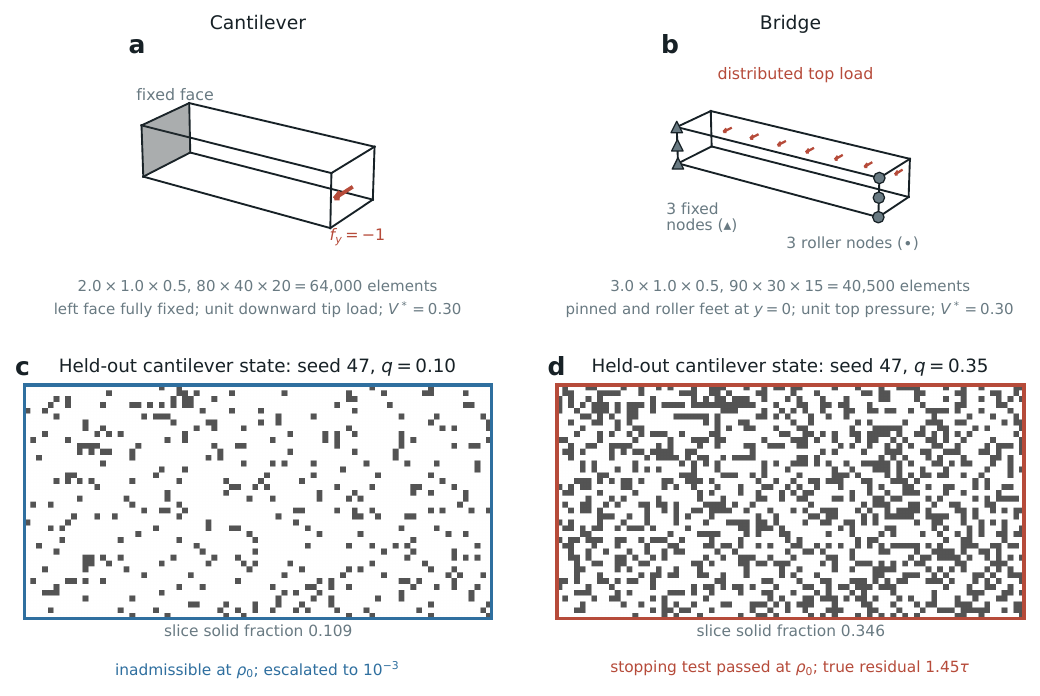}
  \caption{The two problem families that carry the primary evidence, and what a frozen state looks like. (a) Cantilever: \(2.0\times1.0\times0.5\) discretized by \(80\times40\times20\) elements, left face fully fixed, unit downward point load at the centre of the free end. (b) Bridge: \(3.0\times1.0\times0.5\) discretized by \(90\times30\times15\) elements, three fully fixed and three roller nodes on the lower edges, unit pressure on the top face. Both use \(E_0=1\), \(\nu=0.3\) and volume fraction \(0.30\). (c,~d) Mid-depth slices (layer 10 of 20; solid elements carry \(\bar{\rho}_e=1\), void elements \(\bar{\rho}_e=\rhoMin\)) of two held-out Bernoulli states on the cantilever mesh: at \(q=0.10\) the original floor is inadmissible and the state is escalated to \(10^{-3}\); at \(q=0.35\) the projected-residual stopping test passes at the original floor although the recomputed residual is \(1.45\times\) the tolerance (Figure~\ref{fig:policy-guard}b). Frame colours follow the paper's convention: blue, escalated to \(10^{-3}\); red, stopping test passed but recomputed residual failed.}
  \label{fig:setup}
\end{figure}

\paragraph{Benchmark geometries.}
Table~\ref{tab:geometries} specifies every domain used in the paper, and Figure~\ref{fig:setup} draws the two that carry the primary evidence together with two of the frozen states solved on them. The random stress-test states are Bernoulli fields on these same meshes: for solid probability \(q\) and seed \(s\), a field \(u\sim\mathrm{Uniform}(0,1)^{n_e}\) is drawn with NumPy \texttt{default\_rng(s)} and elements with \(u_e<q\) are solid, so \(q\) controls how much of the domain is at the floor.

\begin{table}[!tbp]
\centering
\caption{Domains, meshes, and boundary conditions for every geometry used in the paper. Coordinates are \((x,y,z)\) with the origin at the lower left front corner; loads act in \(-y\) unless stated. All elements are geometrically cubic.}
\label{tab:geometries}
\scriptsize
\setlength{\tabcolsep}{3pt}
\begin{tabularx}{\linewidth}{@{}p{0.095\linewidth}p{0.105\linewidth}p{0.21\linewidth}rp{0.155\linewidth}Y@{}}
\toprule
Family & Domain & Mesh & \(V^{*}\) & Supports & Load \\
\midrule
Cantilever & \(2.0\times1.0\times0.5\) & \(80\times40\times20\) (64k); \(120\times60\times30\) (216k); \(160\times80\times40\) (512k) & 0.30 & \(x=0\) face fully fixed & Unit point load at \((2.0,0.5,0.25)\) \\
Bridge & \(3.0\times1.0\times0.5\) & \(90\times30\times15\) (40.5k); \(210\times70\times35\) (514.5k) & 0.30 & Fully fixed at \((0,0,z)\) and rollers free in \(x\) at \((3,0,z)\), \(z\in\{0,0.25,0.5\}\) & Unit pressure on the \(y=L_y\) face \\
Bracket & \(1.0\times2.0\times0.5\) & \(80\times160\times40\) (512k) & 0.30 & \(y=L_y\) face fully fixed & Oblique point load \((f_x,f_y)=(0.3,-0.7)\) at \((1.0,0,0.25)\) \\
MBB beam & \(3.0\times1.0\times0.5\) & \(210\times70\times35\) (514.5k) & 0.50 & \(x=0\) face pinned in \(x\); \((3,0,0.25)\) pinned in \(y\) & Unit point load at \((0,1.0,0.25)\) \\
Torsion & \(3.0\times1.0\times1.0\) & \(165\times55\times55\) (499k) & 0.25 & \(x=0\) face fully fixed & Opposed end loads \((0,\mp0.5,\pm0.5)\) at \((3,1,0.5)\) and \((3,0,0.5)\) \\
Column & \(1.0\times4.0\times1.0\) & \(50\times200\times50\) (500k) & 0.20 & \(y=0\) face fully fixed & Unit pressure on the \(y=L_y\) face \\
Reduced atlas & \(2.0\times1.0\times0.5\) & \(24\times12\times6\) & --- & \(x=0\) face fully fixed & Unit point load at \((2.0,0.5,0.25)\) \\
\bottomrule
\end{tabularx}
\end{table}

\subsection{Frozen state, floor scaling, and the admissibility target}

For one optimization iteration, let \(\widehat{\rho}_e\in[0,1]\) be the stored, floor-independent physical density of element \(e\), and let \(\bar{\rho}_e(\rhoMin)\) be the density used inside the stiffness interpolation for a candidate floor. For optimized-density transfer states, \(\bar{\rho}_e(\rhoMin)=\widehat{\rho}_e\). For Bernoulli hard-field stress tests the binary support is fixed by a random seed and solid probability \(q\); solid elements use \(\bar{\rho}_e=1\) and void elements use \(\bar{\rho}_e(\rhoMin)=\rhoMin\). With SIMP exponent \(p_{\mathrm{SIMP}}\),
\begin{equation}
  E_e(\rhoMin)=\rhoMin+(1-\rhoMin)\bar{\rho}_e(\rhoMin)^{p_{\mathrm{SIMP}}},
  \label{eq:floor_scaling}
\end{equation}
and the constrained free-DOF operator for the frozen state is
\begin{equation}
  \bK(\rhoMin)=\sum_e E_e(\rhoMin)\bK_e ,
  \label{eq:frozen_operator}
\end{equation}
with \(\bK_e\) the trilinear hexahedral element contribution after boundary-condition restriction. Frozen-state experiments use \(p_{\mathrm{SIMP}}=4.5\) except in the explicit exponent sweep. The solver target is the relative residual
\begin{equation}
  \eta_k(\rhoMin)=\frac{\|\bff-\bK(\rhoMin)\bu_k\|_2}{\|\bff\|_2}.
  \label{eq:relative_residual}
\end{equation}

\paragraph{Operational definition.}
For a fixed density field, hierarchy, preconditioned Krylov method, tolerance \(\tau\), and budget \(k_{\max}\), a floor \(\rhoMin\) is \emph{solver-admissible} if the solve attains \(\eta_k(\rhoMin)\le\tau\) for some \(k\le k_{\max}\). The definition is deliberately operational: a positive floor may make the constrained matrix nonsingular and still be inadmissible for a particular hierarchy at the requested tolerance. Throughout, \(\rho_0=10^{-12}\), the escalation ladder is \(\mathcal{L}=\{10^{-3},10^{-2}\}\), \(\tau=10^{-6}\), and \(k_{\max}=300\).

\paragraph{Which residual counts.}
The \emph{acceptance} residual is the true unpreconditioned free-DOF residual, recomputed with the same matrix-free operator path used by the solve, after the solve returns; constrained degrees of freedom are removed before the norm is formed, and the evaluation uses the FP64 path. The probe features \(r_{50}\), \(r_{100}\), and the plotted iteration histories are the recorded FGMRES residual-history values: after each cycle-start true residual, within-cycle entries are Givens/Hessenberg projected estimates for right-preconditioned FGMRES. They are classifier and visualization features only. No result in this paper accepts a solution on the basis of a projected estimate or of the solver's internal convergence flag. Table~\ref{tab:residual-semantics} states the three quantities side by side.

\begin{table}[!tbp]
\centering
\caption{Residual quantities and their roles. Only the recomputed true residual is ever used to accept a solution.}
\label{tab:residual-semantics}
\scriptsize
\setlength{\tabcolsep}{3pt}
\begin{tabularx}{0.98\linewidth}{@{}p{0.20\linewidth}p{0.34\linewidth}Y@{}}
\toprule
Quantity & What it represents & Role \\
\midrule
Probe features \(r_{50}\), \(r_{100}\) & Recorded FGMRES residual-history values. After each cycle-start true residual, within-cycle entries are Givens/Hessenberg projected estimates for right-preconditioned FGMRES; the scalars are accumulated in FP64 while operator applications use the configured matrix-free path. With restart equal to the 300-iteration cap, no restart boundary is crossed before stopping. & Classifier features only. \\
Internal convergence flag & Set when the projected estimate reaches the tolerance inside the FGMRES cycle; not a fresh evaluation of \(\bff-\bK\bu\). & Diagnostic only; never used to accept. \\
Recomputed true residual & \(\|\bff-\bK(\rhoMin)\bu\|/\|\bff\|\) formed from the matrix-free operator after the selected solve, at the selected floor, on free degrees of freedom, in FP64. & The sole acceptance criterion. \\
\bottomrule
\end{tabularx}
\end{table}

\subsection{What floor escalation can and cannot guarantee}

\begin{proposition}[Monotone floor regularization for fixed densities]
Assume each element stiffness contribution is positive semidefinite, the fully solid constrained stiffness matrix is positive definite on the free degrees of freedom, and \(\bar{\rho}_e\) is held fixed while the floor changes. If \(0<\rho_a<\rho_b\le1\), then
\begin{equation}
  \bK(\rho_b)-\bK(\rho_a)=(\rho_b-\rho_a)\sum_e(1-\bar{\rho}_e^{p_{\mathrm{SIMP}}})\bK_e\succeq 0,
  \label{eq:floor_monotonicity}
\end{equation}
and \(\bK(\rhoMin)\succeq\rhoMin\bK_{\mathrm{solid}}\), so the lower coercivity bound improves at least linearly with \(\rhoMin\).
\end{proposition}

\begin{remark}
Equation~\eqref{eq:floor_monotonicity} applies directly to optimized-density transfer states, where the stored field is fixed. The Bernoulli stress tests are monotone by a one-element argument: void elements use the scale \(\rhoMin+(1-\rhoMin)\rhoMin^{p_{\mathrm{SIMP}}}\), whose derivative in \(\rhoMin\) is \(1-\rhoMin^{p_{\mathrm{SIMP}}}+p_{\mathrm{SIMP}}(1-\rhoMin)\rhoMin^{p_{\mathrm{SIMP}}-1}\ge 0\) on \(0<\rhoMin\le1\). None of this proves multigrid convergence, and the coercivity bound does not imply monotone conditioning improvement. It justifies only why escalation can regularize the operator; whether the reported hierarchy, smoother, coarse correction, and restart policy converge is decided empirically and verified by the acceptance guard. When the ladder returns \(10^{-3}\) or \(10^{-2}\), the claim is \emph{the smallest tested floor in \(\mathcal{L}\)}, not the minimal admissible floor.
\end{remark}

\subsection{Probe features and the preserve/escalate rule}

The detector uses a 100-iteration baseline probe at \(\rho_0\). If the probe converges before a checkpoint, the recorded feature is the last available value, so
\begin{equation}
  r_{50}=\eta_{\min(50,k_{\mathrm{stop}})}(\rho_0),\qquad
  r_{100}=\eta_{\min(100,k_{\mathrm{stop}})}(\rho_0),
\end{equation}
together with the plateau ratio \(r_{100}/r_{50}\). These separate two observed failure modes: states that are immediately bad, where the residual is still large at iteration 50, and states that descend and then stall above tolerance. The rule escalates the floor if
\begin{equation}
  r_{50}\ge10^{-2}
  \quad\text{or}\quad
  \bigl(r_{100}\ge10^{-4}\ \text{and}\ r_{100}/r_{50}\ge0.6\bigr).
  \label{eq:trigger}
\end{equation}
Thresholds were fixed on 16 threshold-development states before the held-out audit and were not retuned; Section~\ref{sec:robust} reports their sensitivity.

\subsection{Guarded policy}\label{sec:guard}

Algorithm~\ref{alg:floor_policy} is a guarded policy, not a standalone classifier. The probe chooses the initial branch and can avoid the cost of a full failed original-floor solve; the selected solve is always run independently, and the probe iterate is never reused as the accepted solution. A predicted preserve means ``try the original floor first,'' not ``accept the original floor.'' Figure~\ref{fig:policy-guard}a shows the control flow and Table~\ref{tab:solver-config} the settings.

\begin{algorithm}[!tbp]
\caption{Guarded residual-probe floor selection}
\label{alg:floor_policy}
\begin{algorithmic}[1]
\Require density field \(\rr\), original floor \(\rho_0\), ladder \(\mathcal{L}\), tolerance \(\tau\), budget \(k_{\max}\)
\State Probe: run GMG-FGMRES at \(\rho_0\) for at most 100 iterations; record \(r_{50}\), \(r_{100}\).
\State \(\mathcal{C}\gets\mathcal{L}\) if the rule \eqref{eq:trigger} fires, else \((\rho_0,\mathcal{L})\) in order.
\For{\(\rhoMin\in\mathcal{C}\)}
  \State Solve the frozen state at \(\rhoMin\) with budget \(k_{\max}\).
  \State Recompute the true relative residual \(\eta=\|\bff-\bK(\rhoMin)\bu\|/\|\bff\|\). \Comment{acceptance guard}
  \If{\(\eta\le\tau\)} \Return floor \(\rhoMin\) and solution \(\bu\). \EndIf
\EndFor
\State \Return no admissible floor in the tested ladder.
\end{algorithmic}
\end{algorithm}

\begin{table}[!tbp]
\centering
\caption{Solver and policy settings used in all reported experiments.}
\label{tab:solver-config}
\scriptsize
\setlength{\tabcolsep}{4pt}
\begin{tabularx}{0.96\linewidth}{p{0.26\linewidth}Y}
\toprule
Component & Setting \\
\midrule
Material and floors & \(p_{\mathrm{SIMP}}=4.5\); original floor \(\rho_0=10^{-12}\); ladder \(\{10^{-3},10^{-2}\}\). \\
Probe and rule & 100 FGMRES iterations at \(\rho_0\); escalate if \(r_{50}\ge10^{-2}\) or (\(r_{100}\ge10^{-4}\) and \(r_{100}/r_{50}\ge0.6\)). \\
Acceptance & Recomputed true relative residual \(\le10^{-6}\) within 300 FGMRES iterations; failed preserve attempts fall back to the ladder. \\
Matrix-free stack & Fine matrix never assembled; four-level geometric multigrid preconditioner, 2:1 trilinear transfer, Galerkin coarse operators, Chebyshev smoothing, FP64 fine path. \\
Coarse correction & Coarse-grid correction scaled by a residual line search and rejected when it fails an acceptance ratio; 10 inner Krylov steps with restart 10 on the coarse level. \\
Fine-level adaptive correction & An additional correction on the finest (root) level: 20 steps of an inner FGMRES iteration with restart 20, preconditioned by a matrix-free \(3\times3\) node-block Jacobi operator (one dense block per mesh node, coupling \(u_x,u_y,u_z\)), applied only if it reduces the residual norm. \\
Outer Krylov & Right-preconditioned FGMRES, 300-iteration cap, restart 300. \\
Connectivity & Binary support diagnostic at density threshold 0.5; reported, never used as a policy input. \\
\bottomrule
\end{tabularx}
\end{table}

\paragraph{Why a flexible Krylov method on a symmetric operator.}
The free-DOF operator \(\bK(\rhoMin)\) of Eq.~\eqref{eq:frozen_operator} is symmetric positive definite, so conjugate gradients with a symmetric fixed preconditioner would normally be preferred \citep{amir2014multigrid}. Here, however, the coarse-grid correction is scaled by a residual-dependent line search and may be rejected, while the additional fine-level correction is an inner FGMRES solve applied only when it reduces the residual norm. The resulting preconditioner is an iterate-dependent nonlinear map, outside the standard assumptions of preconditioned CG; we therefore use right-preconditioned FGMRES \citep{saad1993fgmres}. Flexibility alone does not explain the acceptance failure. The implementation stops on an inexpensive projected residual estimate, which can drift from a freshly evaluated \(\|\bff-\bK\bu\|\) in finite precision; Section~\ref{sec:control} tests this attribution directly with the adaptive components disabled. The guard of Section~\ref{sec:guard} removes dependence on that estimate at acceptance and is compatible with other Krylov schemes.

The values \(p_{\mathrm{SIMP}}=4.5\) and \(\rho_0=10^{-12}\) are stress settings used to expose low-floor inadmissibility, not recommended defaults. Section~\ref{sec:robust} checks \(p_{\mathrm{SIMP}}\in\{3,4,4.5\}\) and \(\rho_0\in\{10^{-12},10^{-9},10^{-8},10^{-6}\}\), including the less stringent \(\rho_0=10^{-6}\) case.

\section{Results: failure, safeguard, and tradeoff}\label{sec:results}

Experiments run on a single NVIDIA GeForce RTX~4090 workstation; the environment is specified in Appendix~\ref{app:repro} and the case families are listed in Table~\ref{tab:case-taxonomy}. The primary evaluation uses 102 held-out random states on the two geometries of Figure~\ref{fig:setup}: cantilever seeds \(41,43,47,53,59,61,67,71\) with \(q\in\{0.08,\dots,0.35\}\) and bridge seeds \(41,43,47,53,59\) with \(q\in\{0.10,\dots,0.35\}\), none used to set thresholds. These Bernoulli fields are deliberately harsher than the smooth, filtered iterates an optimizer produces; the trajectory runs of Section~\ref{sec:in-loop} and the optimizer-generated states of Section~\ref{sec:transfer} test whether the same failure and safeguard occur inside the intended workflow. The evaluation sets nest as follows: the 12-state sensitivity subset of Section~\ref{sec:robust}, and within it the six-state robustness subset of Section~\ref{sec:replication}, are drawn from the 102 held-out states, while the 16 threshold-development states and the 19-state stage-cost set are disjoint from them (Table~\ref{tab:case-taxonomy}).

The results test three linked claims. First, the stopping estimate can fail as an acceptance certificate, and the failure can propagate silently through optimization (Section~\ref{sec:false-keeps}--\ref{sec:in-loop}); a reduced direct diagnostic then isolates the floor--conditioning relation without claiming to explain the full hierarchy (Section~\ref{sec:conditioning}). Second, the residual probe can triage floor selection, but only the recomputed-residual guard controls acceptance (Sections~\ref{sec:heldout} and \ref{sec:cost}). Third, preserving the original operator has a measurable fidelity benefit whose magnitude depends strongly on the state family (Sections~\ref{sec:perturb} and \ref{sec:transfer}). Robustness checks bound, rather than universalize, these claims.

\subsection{Failure of solver-reported convergence}

\subsubsection{Frozen-state false acceptance}\label{sec:false-keeps}

\begin{figure}[!tbp]
  \centering
  \includegraphics[width=\linewidth]{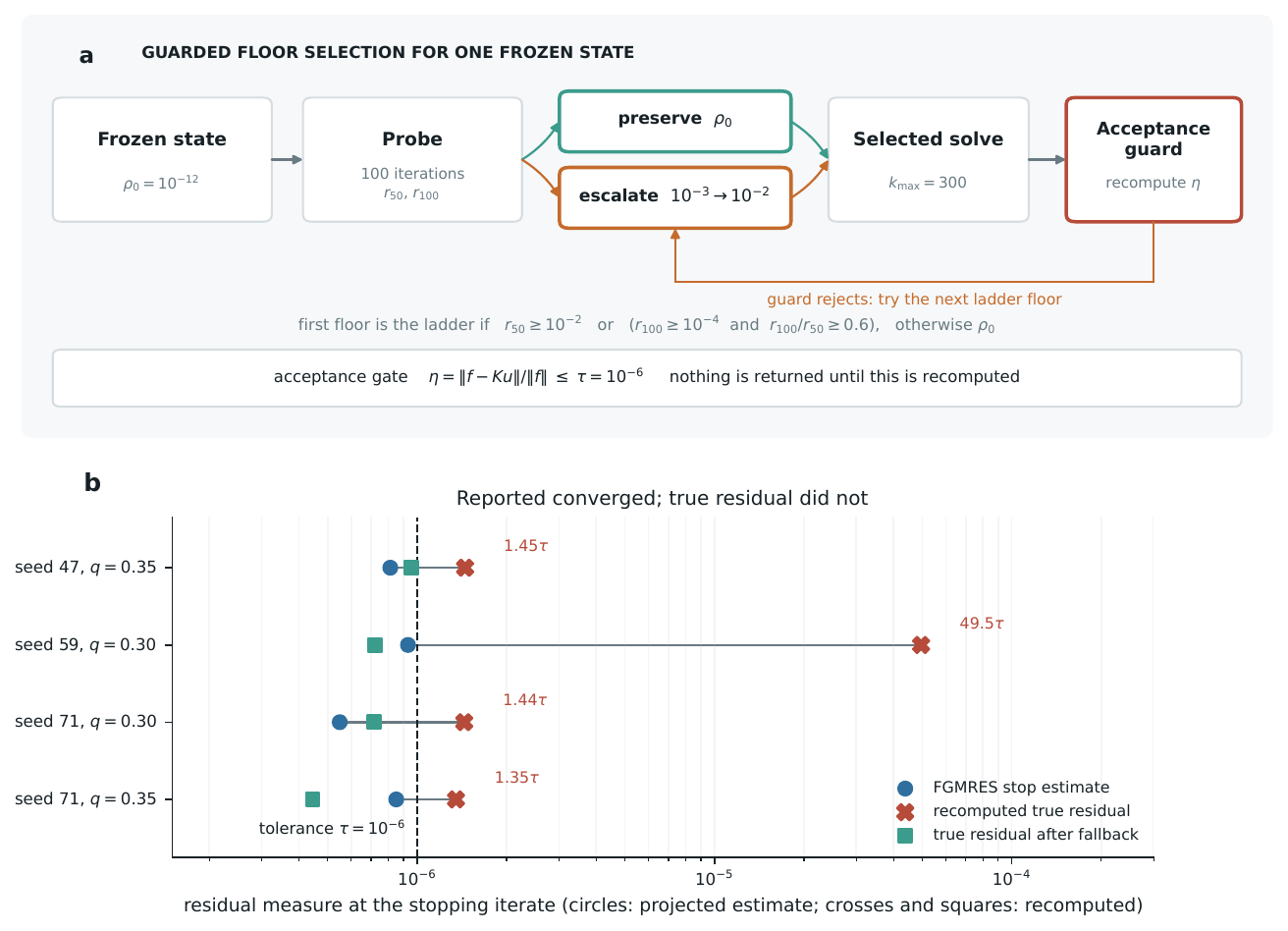}
  \caption{The guarded policy and the failure it exists to catch. (a) Control flow: a 100-iteration probe at \(\rho_0=10^{-12}\) selects the first floor to try, an independent selected solve is run at that floor, and the solution is accepted only if the recomputed true relative residual meets \(\tau=10^{-6}\); otherwise the ladder \(10^{-3}\rightarrow10^{-2}\) is tried in order. (b) The four held-out states on which FGMRES reported convergence at the original floor: the projected stop estimate (circle) is inside tolerance, the recomputed true residual (cross) is not, by factors of \(1.35\times\) to \(49.5\times\), and the fallback solve at \(10^{-3}\) (square) is accepted. Without the guard, all four would be accepted as converged.}
  \label{fig:policy-guard}
\end{figure}

In a strict replay of the 102 held-out states at the original floor, four cantilever states returned with the internal FGMRES convergence flag set and a recomputed true residual above tolerance (Figure~\ref{fig:policy-guard}b, Table~\ref{tab:false-keep-diagnostics}). The projected residual estimate at the stopping iterate was between \(5.49\times10^{-7}\) and \(9.31\times10^{-7}\) in all four cases, that is, comfortably inside the requested \(10^{-6}\); the recomputed residual was \(1.35\times10^{-6}\), \(1.44\times10^{-6}\), \(1.45\times10^{-6}\), and \(4.95\times10^{-5}\), the last exceeding the tolerance by \(49.5\times\). Escalating to \(10^{-3}\) produced accepted solves in 27--34 iterations with true residuals of \(4.45\times10^{-7}\) to \(9.57\times10^{-7}\).

The discrepancy is a projected-estimate versus recomputed-residual stopping mismatch. It is not a preconditioned-residual acceptance criterion, since the acceptance quantity is formed from \(\bff-\bK\bu\) directly, and not a stale restart record, since the outer restart equals the 300-iteration cap and no restart boundary is crossed before stopping (Table~\ref{tab:residual-semantics}). We did not record periodic true-residual checkpoints during the probe, so a checkpointed detector variant remains untested. A workflow that accepts this implementation's stopping flag alone can therefore accept iterates whose recomputed equilibrium residual exceeds the requested tolerance.

\begin{table}[!tbp]
\centering
\caption{The four false-acceptance cases. ``Stop estimate'' is the last recorded FGMRES residual-history value in the strict original-floor replay, at which the solver set its convergence flag; ``true residual'' is the recomputed \(\|\bff-\bK\bu\|/\|\bff\|\) at that iterate. Fallback columns are the guarded selected solve at \(10^{-3}\).}
\label{tab:false-keep-diagnostics}
\scriptsize
\setlength{\tabcolsep}{3.5pt}
\begin{tabularx}{0.98\linewidth}{@{}rrrrrrrrY@{}}
\toprule
Seed & \(q\) & \(r_{50}\) & \(r_{100}\) & Stop estimate & True residual & Orig. iters & Fallback residual & Fallback iters \\
\midrule
47 & 0.35 & \(1.19{\times}10^{-5}\) & \(8.19{\times}10^{-7}\) & \(8.13{\times}10^{-7}\) & \(1.45{\times}10^{-6}\) & 85 & \(9.57{\times}10^{-7}\) & 28 \\
59 & 0.30 & \(8.82{\times}10^{-3}\) & \(6.06{\times}10^{-6}\) & \(9.31{\times}10^{-7}\) & \(4.95{\times}10^{-5}\) & 130 & \(7.23{\times}10^{-7}\) & 34 \\
71 & 0.30 & \(4.54{\times}10^{-4}\) & \(5.54{\times}10^{-7}\) & \(5.49{\times}10^{-7}\) & \(1.44{\times}10^{-6}\) & 100 & \(7.16{\times}10^{-7}\) & 33 \\
71 & 0.35 & \(1.06{\times}10^{-5}\) & \(8.47{\times}10^{-7}\) & \(8.50{\times}10^{-7}\) & \(1.35{\times}10^{-6}\) & 71 & \(4.45{\times}10^{-7}\) & 27 \\
\bottomrule
\end{tabularx}
\end{table}

\subsubsection{Optimization-loop consequence}\label{sec:in-loop}

Frozen states make the mechanism measurable, but the workflow consequence appears only in the loop. We ran 40-iteration optimizations of a 64k-element cantilever and a 40.5k-element bridge under the two-stage continuation schedule of Table~\ref{tab:trajectory-settings} in three configurations: fixed floors \(10^{-12}\), \(10^{-3}\), \(10^{-2}\) with no acceptance test, and the guarded policy applied at every outer iteration.

\begin{figure}[!tbp]
  \centering
  \includegraphics[width=\linewidth]{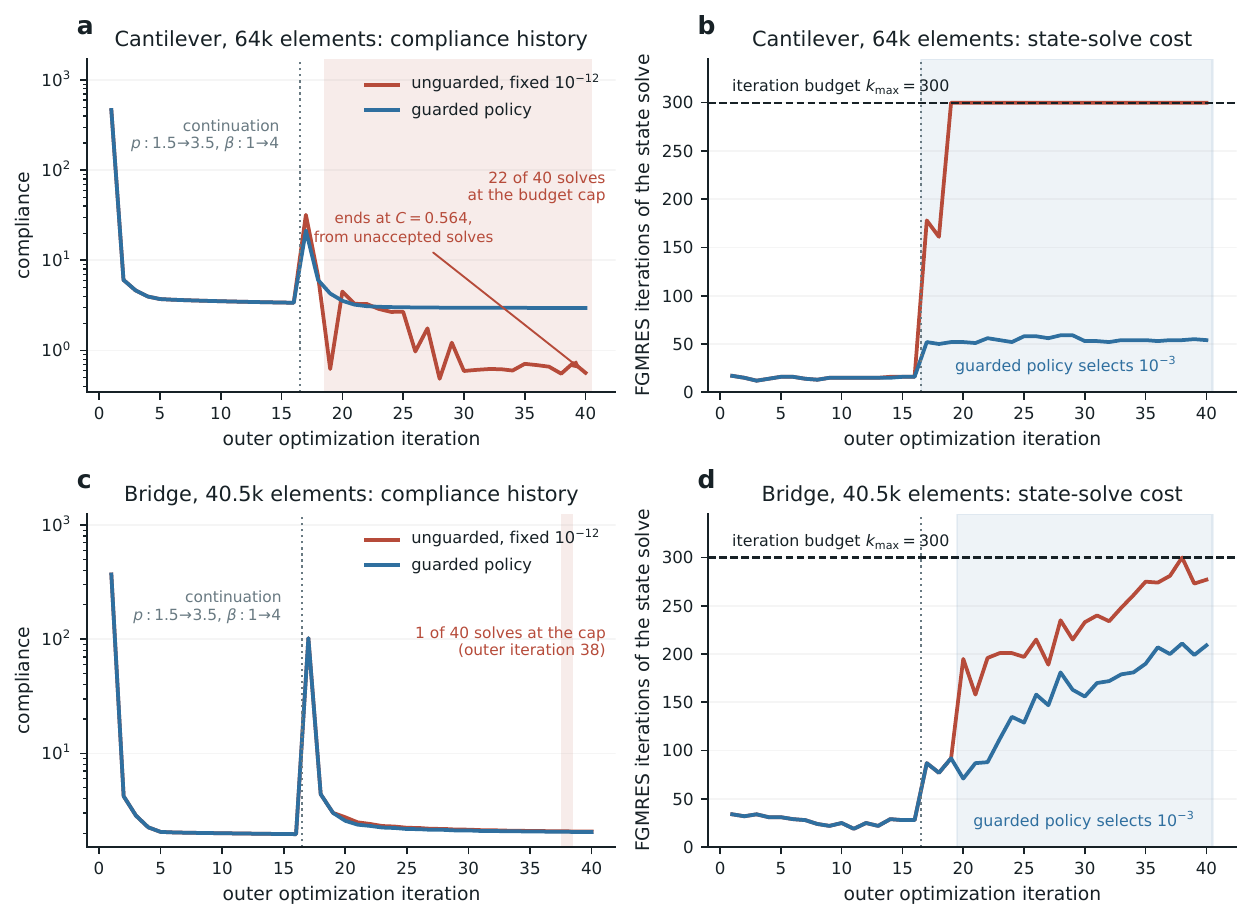}
  \caption{The acceptance failure inside a 40-iteration optimization loop. (a,~b) Cantilever, 64k elements; (c,~d) bridge, 40.5k elements. Left: compliance history for the unguarded fixed-\(10^{-12}\) run and for the guarded policy. Right: FGMRES iterations of the state solve, with the \(k_{\max}=300\) budget marked. The dotted line marks the continuation step (\(p_{\mathrm{SIMP}}:1.5\rightarrow3.5\), \(\beta:1\rightarrow4\)) after outer iteration 15. The unguarded cantilever run saturates the budget from iteration 19 onward and its compliance oscillates between 0.48 and 4.50 with no error raised; the guarded run switches to \(10^{-3}\) at iteration 17 and accepts every solve, the largest accepted true residual being \(9.84\times10^{-7}\). Red shading in the left panels marks the outer iterations whose unguarded state solve returned at the budget cap, so the compliance drawn there comes from unaccepted solves; blue shading in the right panels marks the iterations at which the guarded policy selected \(10^{-3}\).}
  \label{fig:in-loop}
\end{figure}

Figure~\ref{fig:in-loop} shows what the unguarded original-floor run does. Through the first 15 iterations at \(p_{\mathrm{SIMP}}=1.5\) the state solves are easy, requiring 12--17 FGMRES iterations. At outer iteration 16 the continuation step raises \(p_{\mathrm{SIMP}}\) to 3.5 and \(\beta\) to 4, and the state solve immediately deteriorates: from iteration 19 to iteration 40 every solve returns at the 300-iteration cap, and the reported compliance oscillates between \(0.484\) and \(4.502\), ending at \(0.564\). The run reports no failure. Taken at face value, that final value is five times better than any other configuration of the same problem, which is the signature of compliance evaluated on states that never reached equilibrium, not of a better design.

The guarded run on the same case behaves differently at precisely the same point. It keeps the original floor for the first 16 outer iterations, where the probe converges and the guard passes, and switches to \(10^{-3}\) at iteration 17 --- the first iteration after continuation --- on the high-\(r_{50}\) trigger, with \(r_{50}\) jumping from \(6.21\times10^{-7}\) to \(2.54\times10^{-1}\). Every one of its 40 accepted solves satisfies the recomputed residual test, the largest accepted value being \(9.84\times10^{-7}\), and the state solve stays near 52 iterations. The bridge case behaves the same way with a later switch, at iteration 20. Table~\ref{tab:trajectories} collects the six fixed-floor controls and the two guarded runs.

Archived results from two prior 120-iteration optimization runs show the same diagnostic pattern: every state solve terminates at the iteration cap, and the recorded final compliance differs from the last logged iterate, by \(8\%\) for the 514,500-element MBB beam (Section~\ref{sec:transfer}). These states are used as solver stress inputs, not as evidence about optimal designs.

What these runs establish is that the acceptance failure of Section~\ref{sec:false-keeps} is reachable by a standard two-stage continuation schedule, that it is silent, and that the guarded policy converts it into accepted solves at the iteration where it appears. What they do not establish is any claim about final topologies: two guarded trajectories at one resolution cannot support statements about optimization paths, and the fixed-floor controls carry no acceptance record at all, which is why their compliance values are reported as diagnostics rather than as design outcomes.

\begin{table}[!tbp]
\centering
\caption{Forty-iteration trajectories. The fixed-floor controls apply no acceptance test, so their compliance values are diagnostics, not certified design outcomes; the guarded runs accept every solve on a recomputed residual. ``Solves at cap'' counts outer iterations whose state solve returned at \(k_{\max}=300\).}
\label{tab:trajectories}
\scriptsize
\setlength{\tabcolsep}{3.5pt}
\begin{tabularx}{0.98\linewidth}{@{}llrrrrY@{}}
\toprule
Case & Configuration & Final \(C\) & Final grayness & Solves at cap & Wall time (s) & Acceptance record \\
\midrule
Cantilever & fixed \(10^{-12}\) & 0.564 & 0.0595 & 22 of 40 & 312.2 & none \\
Cantilever & fixed \(10^{-3}\) & 2.962 & 0.0353 & 0 & 13.6 & none \\
Cantilever & fixed \(10^{-2}\) & 2.981 & 0.0389 & 0 & 12.6 & none \\
Cantilever & guarded (\(10^{-12}\):16, \(10^{-3}\):24) & 2.962 & 0.0354 & 0 & 227.3 & 40/40 accepted, max \(9.84{\times}10^{-7}\) \\
\midrule
Bridge & fixed \(10^{-12}\) & 2.067 & 0.0733 & 1 of 40 & 148.8 & none \\
Bridge & fixed \(10^{-3}\) & 2.074 & 0.0967 & 0 & 54.7 & none \\
Bridge & fixed \(10^{-2}\) & 1.981 & 0.0574 & 0 & 75.0 & none \\
Bridge & guarded (\(10^{-12}\):19, \(10^{-3}\):21) & 2.049 & 0.0849 & 0 & 233.9 & 40/40 accepted, max \(1.00{\times}10^{-6}\) \\
\bottomrule
\end{tabularx}
\end{table}

\subsection{Floor--conditioning diagnostic}\label{sec:conditioning}

Positive definiteness alone does not quantify the numerical accuracy available at a given floor. We therefore isolate the operator effect on a reduced \(24\times12\times6\) cantilever, for which extreme eigenvalues can be estimated, using 18 fields at eight floors (144 matrices). The measured condition number follows \(\kappa\approx c(\rr)/\rhoMin\), with \(c\) nearly fixed by the field over six decades of \(\rhoMin\). At \(\rho_0=10^{-12}\), \(\kappa\) ranges from \(10^{13}\) to \(1.6\times10^{16}\).

\begin{figure}[!tbp]
  \centering
  \includegraphics[width=0.98\linewidth]{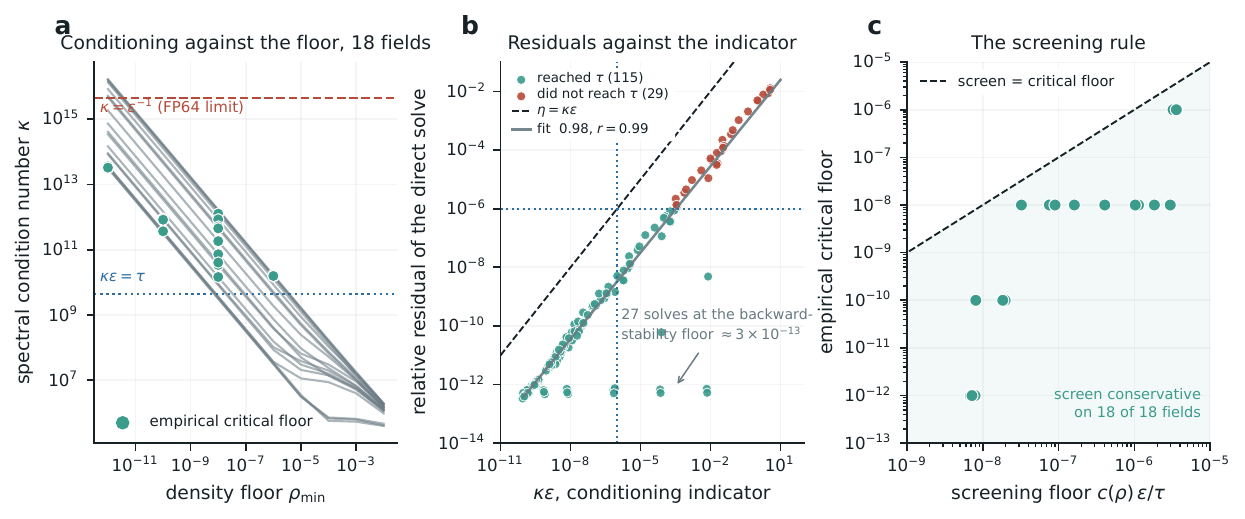}
  \caption{Conditioning of the reduced direct problem against the floor for 18 fields at eight floors each (144 matrices, computed on the CPU with sparse eigensolves; this diagnostic is not used for solve acceptance). (a) Spectral condition number against floor, with \(\kappa=\varepsilon^{-1}\) and \(\kappa\varepsilon=\tau\) marked; teal points are the empirical critical floor of each field, the smallest tested floor at which the direct solve reaches \(\tau\). (b) Attained relative residual against the conditioning indicator \(\kappa\varepsilon\), with the regression quoted in the text; the 27 solves clustered near \(3\times10^{-13}\) sit at the backward-stability floor regardless of \(\kappa\varepsilon\), so the indicator is an upper envelope, loose from below. Every failure lies above \(\kappa\varepsilon=\tau\), while no matrix below that line fails; this empirical separation is not asserted as a universal attainable-residual bound. (c) The screening rule of Eq.~\eqref{eq:screen}: the empirical critical floor of each field against its screen \(c(\rr)\varepsilon/\tau\), with \(c\) taken at the smallest tested floor. All 18 fields lie on or below the diagonal, so the screen is conservative on this set.}
  \label{fig:conditioning}
\end{figure}

The conditioning indicator \(\kappa\varepsilon\) correlates with the attained direct-solve residual: over all 144 solves the log--log slope is \(0.93\) with correlation \(0.91\), and excluding the 27 solves that reach the backward-stability floor near \(3\times10^{-13}\) regardless of \(\kappa\varepsilon\), the slope is \(0.98\) with correlation \(0.99\) (Figure~\ref{fig:conditioning}b). The indicator is an upper envelope for the attained residual, loose from below: all 87 solves with \(\kappa\varepsilon\le\tau\) reach tolerance, whereas 29 of the 57 above that level do not. This empirical separation establishes a floor-dependent finite-precision risk in the reduced operator. It is not used for acceptance and does not locate the admissibility boundary of the four-level GMG--FGMRES solver.

Because \(c\) is fixed by the field to within a few percent over the six smallest tested decades of \(\rhoMin\), the two measurements combine into a screening rule that involves no hierarchy quantity: certification at tolerance \(\tau\) requires \(\kappa\varepsilon\lesssim\tau\), that is,
\begin{equation}
  \rhoMin\gtrsim c(\rr)\,\varepsilon/\tau .
  \label{eq:screen}
\end{equation}
Evaluated with \(c\) taken at the smallest tested floor, the screen ranges from \(6.9\times10^{-9}\) on the easiest of the 18 fields to \(3.6\times10^{-6}\) on the hardest, and it is conservative on all 18: no field's empirical critical floor exceeds its screen (Figure~\ref{fig:conditioning}c). The rule is a necessary-side safeguard only --- 28 of the 57 solves at floors below their screen still converged --- but it explains why a conventional floor of \(10^{-6}\) is marginal for the hardest fields, consistent with the escalations that persist at \(\rho_0=10^{-6}\) in Section~\ref{sec:robust}. It is a statement about the reduced operator under a backward-stable direct solve; the multigrid hierarchy decides where, above this bound, a particular state actually falls.

\subsection{Safeguard performance and preservation cost}

\subsubsection{Held-out reference test}\label{sec:heldout}

\begin{figure}[!tbp]
  \centering
  \includegraphics[width=\linewidth]{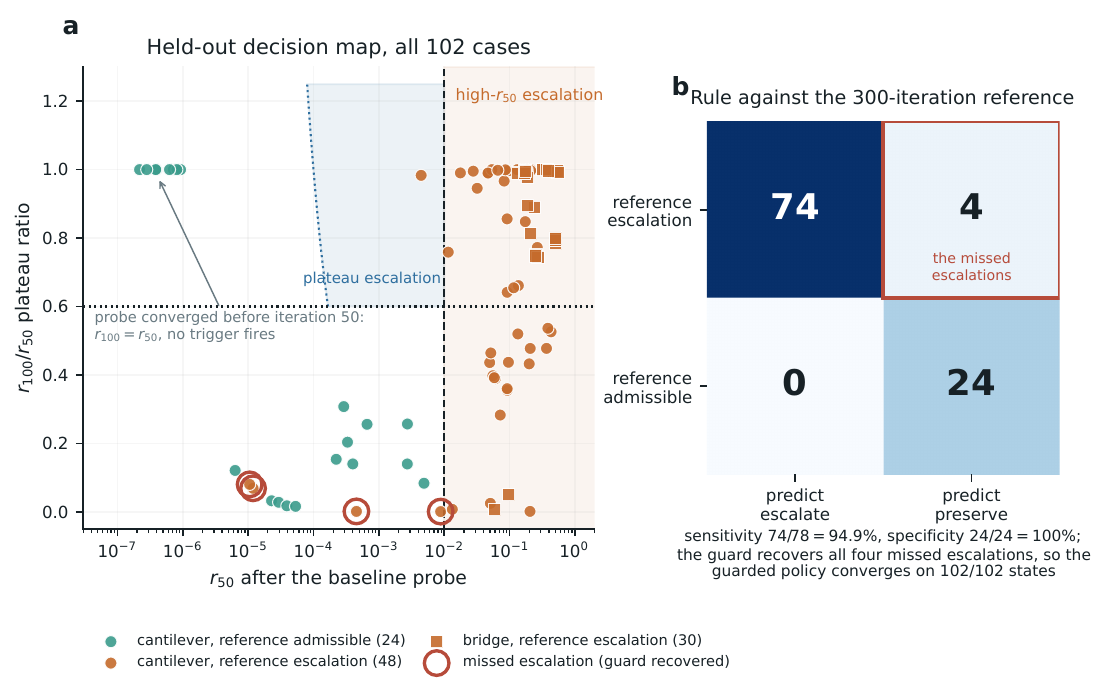}
  \caption{Held-out evaluation on 102 unseen random states. (a) Decision map in the two probe features. Marker shape distinguishes geometry, fill colour gives the reference classification from the 300-iteration original-floor replay, shaded regions are the two escalation triggers, and red rings mark the four missed escalations recovered by the acceptance guard. (b) Confusion matrix of the rule against the reference classification; all 102 selected solves satisfy the recomputed-residual criterion.}
  \label{fig:heldout}
\end{figure}

A 300-iteration replay at the original floor provides reference classifications for the 102 held-out states: 78 require escalation and 24 are admissible. Against those classifications the rule predicts 74 escalations and 28 preserves, giving TP/TN/FP/FN \(=74/24/0/4\) with escalation as the positive class, sensitivity \(74/78=94.9\%\) (Wilson 95\% interval 87.5--98.0\%), and specificity \(24/24=100\%\) (86.2--100\%). The four false negatives are the four missed escalations of Section~\ref{sec:false-keeps}; thus \(4/28=14.3\%\) (5.7--31.5\%) of predicted preserves require guard-triggered fallback. With the guard in place, all 102 selected solves satisfy the recomputed-residual criterion (Wilson 96.4--100\%). The intervals summarize this structured seed/\(q\) test set; they are not population estimates over optimization practice. Figure~\ref{fig:heldout} shows the decision map and confusion matrix, and Table~\ref{tab:heldout} the per-block outcomes.

The bridge block is uniform: all 30 states trigger the high-\(r_{50}\) rule, select \(10^{-3}\), and are accepted. The states at plateau ratio exactly 1 and \(r_{50}<10^{-6}\) are probes that converge before iteration 50, for which \(r_{100}=r_{50}\); they do not trigger the plateau rule because \(r_{100}\) is far below \(10^{-4}\). The four missed escalations are not equally deep inside the preserve region. Three lie at \(r_{50}\le4.54\times10^{-4}\) with plateau ratios below \(0.11\), interleaved with reference-admissible states on both features, so no setting of this two-feature rule separates them. The fourth, seed~59 at \(q=0.30\), lies at \(r_{50}=8.82\times10^{-3}\), within \(12\%\) of the high-\(r_{50}\) threshold, and is also the largest violation observed (\(49.5\tau\)); a threshold of \(7.5\times10^{-3}\) would have flagged it, the development sweep of Figure~\ref{fig:si-threshold} shows that value performing identically to the reported \(10^{-2}\) on the states used to fix it, and the largest reference-admissible \(r_{50}\) in the held-out set is \(4.9\times10^{-3}\), so the tighter threshold would also have produced no false escalation here. We did not retune, because selecting a threshold on held-out outcomes would invalidate the evaluation. The classifier was therefore within reach of the largest violation and out of reach of the three smallest; the recomputed-residual guard, which recovers all four without reference to either threshold, is the load-bearing component.

\begin{table}[!tbp]
\centering
\caption{Guarded held-out outcomes by block. ``Guard fallback'' counts predicted preserves whose original-floor solve failed the recomputed-residual test and was escalated.}
\label{tab:heldout}
\small
\setlength{\tabcolsep}{4pt}
\begin{tabularx}{0.98\linewidth}{@{}p{0.22\linewidth}rrrrrY@{}}
\toprule
Block & States & Pred.\ escalate & Pred.\ preserve & Guard fallback & Converged & Failed \\
\midrule
Cantilever held-out & 72 & 44 & 28 & 4 & 72 & 0 \\
Bridge held-out & 30 & 30 & 0 & 0 & 30 & 0 \\
\textbf{Total} & \textbf{102} & \textbf{74} & \textbf{28} & \textbf{4} & \textbf{102} & \textbf{0} \\
\bottomrule
\end{tabularx}
\end{table}

\paragraph{How much margin do preserved floors have?}
Because admissibility is defined relative to \(k_{\max}\), a preserved-floor solve that consumes most of the budget is fragile. Of the 24 held-out preserves, 21 are accepted within 195 iterations, but three require 246, 246, and 274 of the 300 available; among the optimized-density preserves of Section~\ref{sec:transfer}, the MBB beam needs 247. Four of the 31 preserve decisions in this paper therefore use at least \(80\%\) of the iteration budget, and would flip to escalations under a modest reduction of \(k_{\max}\). We report this rather than treating preserve/escalate as a budget-independent property of the state.

\subsubsection{Cost of preserving the original operator}\label{sec:cost}

\begin{figure}[!tbp]
  \centering
  \includegraphics[width=\linewidth]{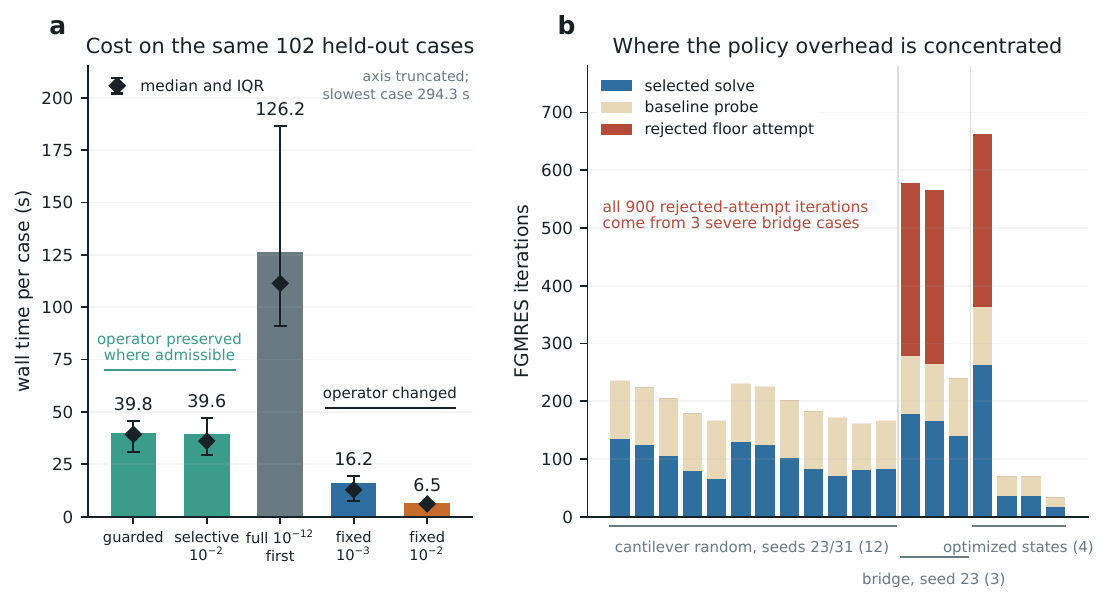}
  \caption{Cost of the guarded policy. (a) Per-case wall time on the 102 held-out states: bars are means, diamonds medians, whiskers the interquartile range. The two fixed-floor policies are faster and converge on every state, but replace the operator on all 102, including the 24 that did not need it. (b) Iteration-equivalent overhead per transfer case, split into selected solve, baseline probe, and rejected floor attempts; all 900 rejected-attempt iterations come from three severe bridge states.}
  \label{fig:cost}
\end{figure}

Table~\ref{tab:timing} and Figure~\ref{fig:cost} give per-case wall times for five policies on the same 102 states. These are single observed timings per case, with probe, setup, selected solve, and fallback included; they are not repeated-run microbenchmarks, and differences of a few percent should not be interpreted.

The guarded policy averages 39.8~s per case. Exhausting the original floor before falling back averages 126.2~s, so the probe removes most of the cost of wasted original-floor attempts. The sharper comparison is with the fixed floors, which also converge on all 102 states: \(10^{-3}\) averages 16.2~s and \(10^{-2}\) averages 6.5~s. The guarded policy is \(2.5\times\) slower than always solving at \(10^{-3}\) and \(6.1\times\) slower than always solving at \(10^{-2}\). What it buys is that on the 24 states where the original floor is admissible, the operator is not modified --- and Section~\ref{sec:perturb} quantifies what modifying it costs. A severity-aware variant that sends predicted escalations with \(r_{50}\ge0.5\) directly to \(10^{-2}\) converges on all 102 states at 39.6~s, indistinguishable from the reported ladder at this measurement precision; it does not reduce the hardware-independent cost either, averaging 105.5 FGMRES iterations against 100.7 for the reported ladder. On this matrix the refinement is safe and not worth its complexity.

Overhead is concentrated rather than spread (Figure~\ref{fig:cost}b). Per-stage accounting --- probe, selected solve, and rejected floor attempts counted separately --- was recorded on a 19-state stage-cost set disjoint from both the development and held-out sets: 12 cantilever random fields at seeds 23 and 31 with \(q\in\{0.10,0.12,0.15,0.18,0.20,0.35\}\), 3 bridge random fields at seed 23 with \(q\in\{0.10,0.20,0.35\}\), and the 64k cantilever, 512k cantilever, 512k bracket, and 514.5k bridge optimized designs. Across those 19 states the mean selected solve costs 106.1 FGMRES iterations and the mean recorded policy cost is 240.4, with a maximum policy-to-selected ratio of 3.42. All 900 rejected-attempt iterations come from three severe states --- bridge \(q=0.10\), bridge \(q=0.20\), and the optimized bridge density --- each of which spends a full 300-iteration budget at \(10^{-3}\) before escalating. That is the concrete target for a better ladder, and it is why we report ladder cost separately from probe cost.

\begin{table}[!tbp]
\centering
\caption{Observed per-case wall time on the 102-state held-out test set (seconds; single observed timings). All five policies converge on all 102 states.}
\label{tab:timing}
\scriptsize
\setlength{\tabcolsep}{4pt}
\begin{tabularx}{0.98\linewidth}{@{}p{0.30\linewidth}rrrrrY@{}}
\toprule
Policy & Mean & Median & IQR low & IQR high & Max & Selected floors \\
\midrule
Guarded residual-probe policy & 39.8 & 39.3 & 30.9 & 45.6 & 106.4 & \(10^{-12}\):24; \(10^{-3}\):78 \\
Selective entry at \(10^{-2}\) & 39.6 & 36.1 & 29.3 & 46.9 & 101.9 & \(10^{-12}\):24; \(10^{-3}\):66; \(10^{-2}\):12 \\
Full \(10^{-12}\) then fallback & 126.2 & 111.4 & 91.2 & 186.4 & 294.3 & \(10^{-12}\):24; \(10^{-3}\):72; \(10^{-2}\):6 \\
Fixed \(10^{-3}\) & 16.2 & 12.8 & 7.7 & 19.5 & 54.2 & \(10^{-3}\):102 \\
Fixed \(10^{-2}\) & 6.5 & 6.1 & 5.3 & 7.2 & 11.9 & \(10^{-2}\):102 \\
\bottomrule
\end{tabularx}
\end{table}

\subsection{Operator fidelity and transfer}

\subsubsection{Perturbation from a fixed raised floor}\label{sec:perturb}

\begin{figure}[!tbp]
  \centering
  \includegraphics[width=\linewidth]{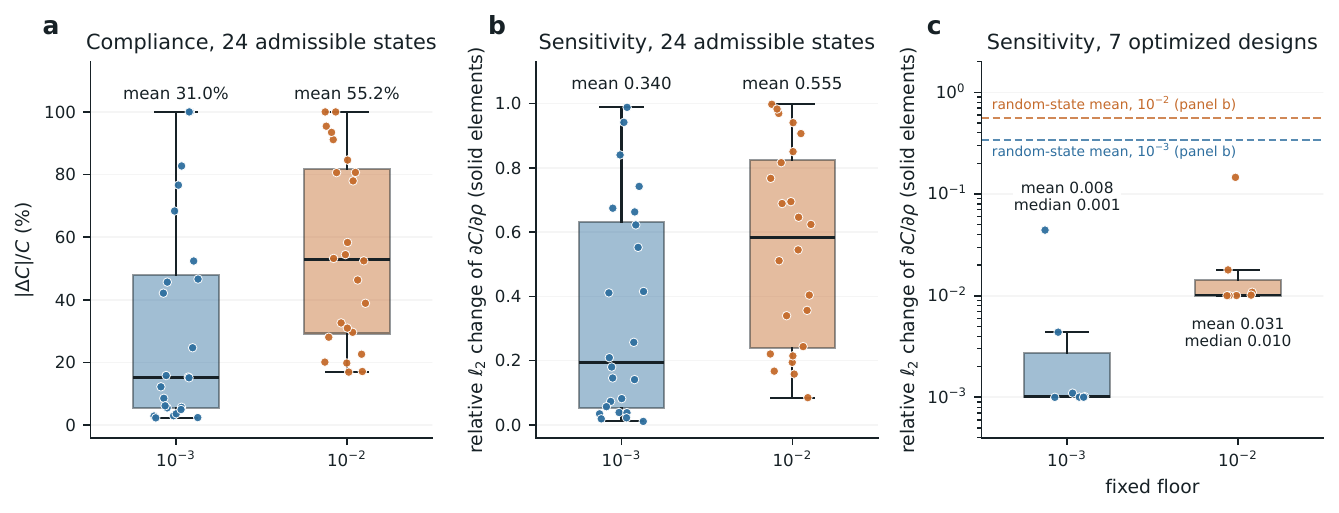}
  \caption{What a fixed raised floor changes on states that did not need it. (a) Absolute relative compliance change and (b) relative \(\ell_2\) change of the solid-element compliance gradient, on the 24 held-out random states where the original floor is admissible; boxes are quartiles, points individual states. Every one of the 24 changes in (a) is a \emph{decrease}, and the two states that reach \(100\%\) are not clipped: their original-floor compliance is of order \(10^{11}\), and raising the floor removes essentially all of it. (c) The same gradient measurement on the seven optimized designs whose original floor is admissible, on a logarithmic axis, with the panel-(b) means marked for comparison; the mean in (c) is pulled up by a single design, so the median is given as well. The perturbation is one to two orders of magnitude smaller on optimized designs than on severe random states, so the neutrality of a fixed floor cannot be assumed from either family alone.}
  \label{fig:perturb}
\end{figure}

The case for preservation rests on what escalation does to states that did not need it (Figure~\ref{fig:perturb}a,b; Table~\ref{tab:perturbation}). On the 24 held-out states where the original floor is admissible, substituting a fixed floor and recomputing compliance and sensitivities with the same floor-substituted density convention gives, over 48 raised-floor solves that all converge: mean absolute compliance change \(31.0\%\) at \(10^{-3}\) and \(55.2\%\) at \(10^{-2}\); mean relative \(\ell_2\) change of the solid-element compliance gradient \(0.340\) and \(0.555\), with maxima \(0.988\) and \(0.998\); and minimum Pearson correlation between the original-floor and raised-floor gradient vectors of \(0.220\) and \(0.104\). At \(10^{-2}\) the gradient direction on the worst of these states is nearly uncorrelated with the one the model prescribes. Every one of the 48 changes is a reduction in compliance, and the two extreme values --- reported as \(100\%\), and equal to \(99.9999999\%\) rather than saturating at a cap --- come from two states whose original-floor compliance is of order \(10^{11}\): at \(\rho_0=10^{-12}\) those fields are so nearly unsupported that almost all of their compliance is an artifact of the floor, and raising it returns values of order \(10^{1}\).

These states are deliberately severe random fields, so the same measurement was repeated on the seven optimized designs of Section~\ref{sec:transfer} whose original floor is admissible, using the stored density field and the standard SIMP derivative rather than the floor-substituted convention. There the perturbation is one to two orders of magnitude smaller: a fixed \(10^{-3}\) floor changes compliance by \(0.48\%\) on average (maximum \(2.2\%\)) and the solid-element gradient by a mean relative \(\ell_2\) norm of \(0.008\) (maximum \(0.044\)), and a fixed \(10^{-2}\) floor changes compliance by \(2.9\%\) on average (maximum \(10.0\%\)) and the gradient by \(0.031\) (maximum \(0.146\)). Gradient \emph{direction} is essentially untouched on these designs: the minimum Pearson correlation with the original-floor gradient is \(0.997\) at \(10^{-3}\) and \(0.995\) at \(10^{-2}\), against \(0.220\) and \(0.104\) on the random states (Figure~\ref{fig:perturb}c, Table~\ref{tab:perturbation}).

The structure of those numbers is informative. For a nearly binary design the raised floor enters the gradient mainly through the \((1-\rhoMin)\) prefactor, which alone predicts a relative \(\ell_2\) change of exactly \(\rhoMin\); four of the seven designs sit at that value to three digits and a fifth lies close to it. The two that exceed it substantially are the ones with real void content in the load path --- the MBB beam, the grayest field in the set, at \(0.018\), and the column, the lowest volume fraction at \(0.20\), at \(0.146\) --- so the excess above \(\rhoMin\) measures how much the escalated floor actually changes the mechanics rather than merely rescaling the sensitivities. One of these seven, the MBB beam, is the partially converged field discussed in Section~\ref{sec:transfer}; dropping it moves the \(10^{-2}\) means from \(2.9\%\) and \(0.031\) to \(1.7\%\) and \(0.033\), so the conclusion does not depend on it.

A fixed raised floor is not exactly neutral. On the seven optimized states its effect is small enough to be an engineering judgement, whereas on the 24 severe random states a fixed \(10^{-2}\) floor changes compliance by \(55.2\%\) on average and yields a worst-case gradient correlation of \(0.104\). The preservation-first policy therefore has its clearest fidelity benefit on hard intermediate states; the present data do not show that it is necessary for every optimized design.

\begin{table}[!tbp]
\centering
\caption{Operator perturbation summary. Compliance changes are absolute relative changes against the original-floor reference, and all of them are decreases; the two random-state maxima printed as \(100\%\) are \(99.9999999\%\), not a cap. Gradient changes are relative \(\ell_2\) norms of \(\partial C/\partial\rho\) over solid elements, with the minimum Pearson correlation against the original-floor gradient in the last column. Random states use the floor-substituted binary convention of Eq.~\eqref{eq:floor_scaling}; optimized designs use the stored density field and the standard SIMP derivative.}
\label{tab:perturbation}
\small
\setlength{\tabcolsep}{4pt}
\begin{tabular}{@{}llrrrlr@{}}
\toprule
State group & Floor & States & Mean \(|\Delta C|/C\) & Max \(|\Delta C|/C\) & Gradient \(\ell_2\) change & Min \(r\) \\
\midrule
Random admissible states & \(10^{-3}\) & 24 & 31.0\% & 100\% & 0.340 (max 0.988) & 0.220 \\
Random admissible states & \(10^{-2}\) & 24 & 55.2\% & 100\% & 0.555 (max 0.998) & 0.104 \\
Optimized designs & \(10^{-3}\) & \phantom{0}7 & 0.48\% & 2.20\% & 0.008 (max 0.044) & 0.997 \\
Optimized designs & \(10^{-2}\) & \phantom{0}7 & 2.92\% & 10.0\% & 0.031 (max 0.146) & 0.995 \\
Random cantilever controls & \(10^{-3}\) & \phantom{0}2 & 2.76\% & 2.79\% & not measured & --- \\
Random cantilever controls & \(10^{-2}\) & \phantom{0}2 & 18.5\% & 18.6\% & not measured & --- \\
\bottomrule
\end{tabular}
\end{table}

\subsubsection{Transfer beyond synthetic stress states}\label{sec:transfer}

\begin{figure}[!tbp]
  \centering
  \includegraphics[width=\linewidth]{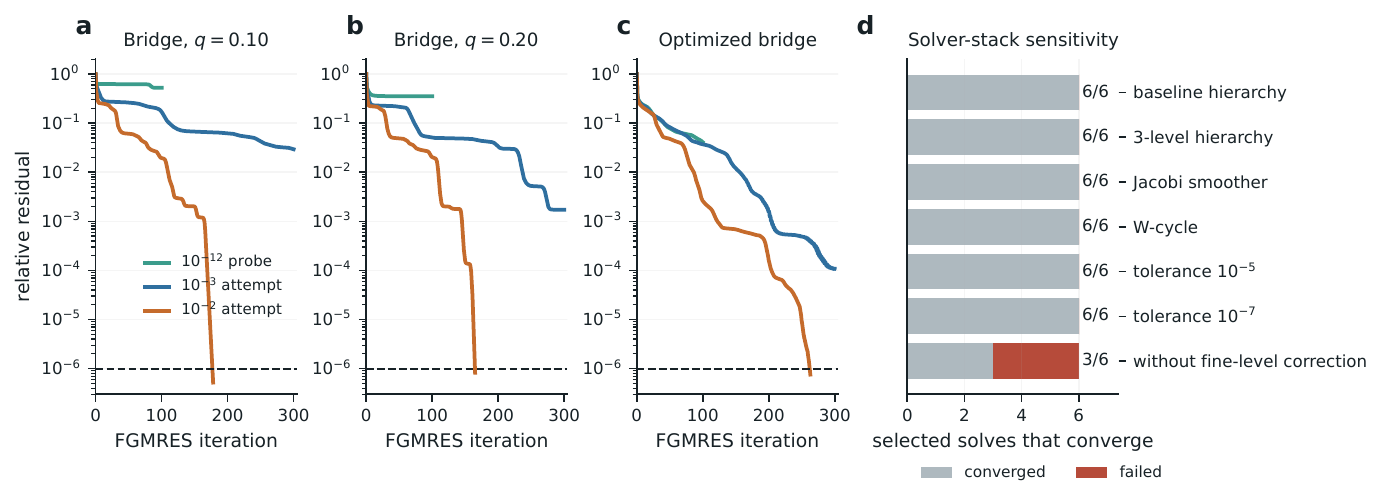}
  \caption{Where escalation is needed, and what the stack contributes. (a--c) Residual histories at the original floor and at each ladder floor for two bridge random states and the optimized bridge density; the dashed line is the \(10^{-6}\) tolerance. In all three the original floor stagnates and \(10^{-3}\) stalls above tolerance. (d) Six-state ablation of the solver stack: every tested variant preserves selected-solve convergence except removing the fine-level adaptive correction --- the node-block inner-FGMRES correction on the finest level defined in Table~\ref{tab:solver-config} --- which converges 3 of 6.}
  \label{fig:rescue}
\end{figure}

Figure~\ref{fig:rescue}a--c shows that escalation is required for two bridge random states and one optimizer-generated bridge state: at the original floor the residual stagnates, at \(10^{-3}\) it descends and stalls above tolerance, and only \(10^{-2}\) reaches \(10^{-6}\). A finer sweep over \(10^{-10},10^{-8},10^{-6},10^{-5},10^{-4},10^{-3},10^{-2}\) on six bridge states selects \(10^{-2}\) four times, \(10^{-3}\) once, and \(10^{-6}\) once. Thus the reported ladder is a conservative discretization of the observed floor boundary; denser ladders trade smaller interventions against additional failed attempts.

Optimized designs matter more than random fields for relevance, and they behave differently from them. Eight unique final density fields from previous matrix-free optimization runs, spanning cantilever, bridge, bracket, MBB, torsion, and column families from 64k to 514.5k elements, were evaluated under the guarded policy (Table~\ref{tab:optimized-transfer}, Figure~\ref{fig:si-gallery}). Seven keep the original floor and are accepted; the 514.5k bridge field triggers the high-\(r_{50}\) rule at \(r_{50}=7.86\times10^{-2}\), does not reach tolerance at \(10^{-3}\), and is accepted at \(10^{-2}\) in 263 iterations. That field is connected under the \(\rho\ge0.5\) support diagnostic, which is why connectivity is reported as a diagnostic and never used as a policy input: on this evidence a connected 3D design can still be solver-inadmissible. This is a single counterexample.

\paragraph{Source-run acceptance limits the interpretation of two transfer states.}
The three cantilevers and bracket have at most \(4\%\) capped source solves and are accepted here in 17--35 iterations. The torsion and column have many capped source solves but internally consistent final metadata and are accepted here in 87 and 158 iterations. For the MBB and bridge fields, every source solve reached its iteration cap and the recorded final metadata disagrees with the last logged iterate; they are therefore treated as optimizer-generated solver stress states, not as converged designs. Full source-run diagnostics are reported in Table~\ref{tab:optimized-source-metadata} and the fields are shown in Figure~\ref{fig:si-gallery}. With only eight states, the observed association between intermediate density and difficulty is not used as a mechanistic or predictive claim.

\paragraph{One-million-element check.}
On a \(10^6\)-element optimized cantilever state in the replication configuration, the policy keeps the original floor and the selected solve is accepted after 50 FGMRES iterations at a recomputed residual of \(6.5\times10^{-7}\). The check establishes that the decision procedure can be executed at this size; it is not a scaling study. The accepted solve uses \(32.3\)~GiB of the \(32.6\)~GiB available, and the unrestarted 300-iteration budget would exceed device memory (Appendix~\ref{app:repro}), so the transfer study stops at 514,500 elements.

\begin{table}[!tbp]
\centering
\caption{Optimized-density transfer. Each family contributes one unique fixed density field; duplicate saved optimizer states have identical density hashes and are not counted as additional samples, but solving them independently gives a view of run-to-run spread: the largest observed difference is 263 versus 262 iterations on the 514.5k bridge, with accepted residuals of \(7.5\times10^{-7}\) and \(1.0\times10^{-6}\). ``Gray'' is the fraction of elements with \(0.05<\widehat{\rho}_e<0.95\). Iterations are for the accepted selected solve at \(\tau=10^{-6}\). The two fields with a non-negligible gray fraction are the two hardest states in the set.}
\label{tab:optimized-transfer}
\small
\setlength{\tabcolsep}{4pt}
\begin{tabular}{@{}llrrrlrr@{}}
\toprule
Family & Geometry & Elements & Gray & \(r_{50}\) & Decision & Selected floor & Iterations \\
\midrule
64k cantilever & cantilever & 64,000 & 0 & \(2.5{\times}10^{-7}\) & preserve & \(10^{-12}\) & \phantom{0}17 \\
216k cantilever & cantilever & 216,000 & \(<\)0.01\% & \(4.3{\times}10^{-7}\) & preserve & \(10^{-12}\) & \phantom{0}27 \\
512k cantilever & cantilever & 512,000 & 0 & \(6.6{\times}10^{-7}\) & preserve & \(10^{-12}\) & \phantom{0}35 \\
512k bracket & bracket & 512,000 & 0 & \(7.7{\times}10^{-7}\) & preserve & \(10^{-12}\) & \phantom{0}35 \\
499k torsion & torsion & 499,125 & 0 & \(9.6{\times}10^{-6}\) & preserve & \(10^{-12}\) & \phantom{0}87 \\
500k column & column & 500,000 & 0 & \(1.5{\times}10^{-5}\) & preserve & \(10^{-12}\) & 158 \\
514.5k MBB beam & MBB & 514,500 & 0.96\% & \(2.0{\times}10^{-3}\) & preserve & \(10^{-12}\) & 247 \\
514.5k bridge & bridge & 514,500 & 0.11\% & \(7.9{\times}10^{-2}\) & escalate & \(10^{-2}\) & 263 \\
\bottomrule
\end{tabular}
\end{table}

\subsection{Robustness and implementation dependence}

\subsubsection{Sensitivity of the decision rule}\label{sec:robust}

Three checks bound how specific the rule is to its own settings. A sweep of 288 threshold combinations over the high-residual threshold, the plateau residual threshold, and the plateau ratio on the 16 development states leaves 160 combinations with no missed escalations, so the reported triplet is not an isolated point (Figure~\ref{fig:si-threshold}). The 12-state post-hoc sensitivity set comprises cantilever and bridge fields at seeds 41 and 43 with \(q\in\{0.10,0.20,0.35\}\); it is a subset of the held-out test set and is not an additional independent test. On this set, \(p_{\mathrm{SIMP}}\in\{3,4,4.5\}\) each give 12 of 12 accepted solves with identical selected-floor counts, and \(\rho_0\in\{10^{-12},10^{-9},10^{-8}\}\) reproduces the same pattern (Figure~\ref{fig:si-sensitivity}).

The original-floor sweep tests whether the result is confined to \(\rho_0=10^{-12}\). At \(\rho_0=10^{-6}\), four of the 12 states keep the original floor and eight require escalation to \(10^{-3}\), with no \(10^{-2}\) selection and all 12 selected solves accepted. This small sensitivity set shows persistence at a less stringent floor; it does not estimate how often escalation is needed in general practice.

A six-state ablation isolates one stack component (Figure~\ref{fig:rescue}d). Three-level and four-level hierarchies, Chebyshev and Jacobi smoothing, V- and W-cycles, and tolerances \(10^{-5}\) and \(10^{-7}\) all converge 6 of 6 with no missed escalations. Removing the fine-level adaptive correction of Table~\ref{tab:solver-config} converges 3 of 6. On this evidence that correction is a material part of where the admissibility boundary sits for this stack, though six states and one disabled component do not identify a mechanism uniquely.

\subsubsection{Cross-configuration and fine-precision replication}\label{sec:replication}

A six-state robustness set is re-run on a second hardware/software configuration and with an FP32 fine-level smoother to test numerical reproducibility.

The replication configuration (Table~\ref{tab:platforms}) differs from the reference configuration in GPU architecture (SM~12.0 versus SM~8.9), CuPy major version (14 versus 13), and CUDA major version (13 versus 12). Re-running the FP64 solver selects the same floor on all six states and returns the same accepted-solve iteration count on five, with the remaining count differing by one iteration (179 versus 180). Probe features agree to at least four significant digits; the largest relative difference in \(r_{50}\) is \(8.0\times10^{-5}\). Wall times are not compared across configurations.

Replacing the FP64 fine-level smoother with an FP32 one --- while retaining FP64 coarse levels, outer operator applications, and acceptance residuals --- gives the same selected floor on all six states and the same iteration count on five (149 versus 150 on the sixth). Probe features agree to \(1.9\times10^{-4}\), and every selected solve is accepted below \(9.35\times10^{-7}\) (Table~\ref{tab:replication}). For these six states, the tested precision change does not move the observed decision boundary. The result does not cover the full held-out test set, other reduced-precision levels, or BF16.

\subsubsection{Fixed-preconditioner control}\label{sec:control}

The acceptance failure of Section~\ref{sec:false-keeps} could in principle be a property of high-contrast SIMP states at this tolerance, or a property of the iterate-dependent preconditioner of Table~\ref{tab:solver-config}. To separate the two, both adaptive components --- the line-searched coarse correction and the conditionally accepted fine-level correction --- were disabled, leaving a fixed linear V-cycle; with a fixed linear preconditioner the flexible outer method coincides with right-preconditioned GMRES. Nine states were re-run in the replication configuration with the escalation rule disabled, so that the original floor is always attempted to exhaustion and every solution is accepted or rejected solely by the recomputed-residual guard: the four false-acceptance states, the two adjacent cantilever states at the same seeds, one reference-admissible cantilever state, and the two severe bridge states of Figure~\ref{fig:rescue}. Table~\ref{tab:control} lists every attempt.

No false acceptance occurred in any of the 18 floor attempts: every stopping flag agreed with the recomputed residual, and every rejected attempt returned with its projected estimate above tolerance, by factors of \(2.8\) to \(10^{6}\), so every failure was visible (Table~\ref{tab:control}). The four false-acceptance states fail visibly at \(10^{-12}\) under the fixed preconditioner and are accepted at \(10^{-3}\) in 15--41 iterations with true residuals of \(4.6\times10^{-7}\) to \(8.2\times10^{-7}\). The control is not a replacement for the reported solver: the reference-admissible cantilever state that the full hierarchy accepts in 30 iterations needs 130, and the two severe bridge states fail at every tested floor with residuals near unity, although the full hierarchy accepts both at \(10^{-2}\). On these states, therefore, the projected-versus-recomputed divergence was observed only with the iterate-dependent preconditioner --- and the components that make the severe states solvable are the same components that make the stopping estimate untrustworthy, which is precisely why the guard is the price of the stronger preconditioner. This is a nine-state control under one alternative configuration, not a general attribution; a symmetric multigrid-preconditioned CG control remains untested.

\begin{table}[!tbp]
\centering
\caption{Six-state robustness results for the reference FP64 configuration, the replication FP64 configuration, and an FP32 fine-level smoother in the replication configuration. Entries are selected floor and accepted-solve iteration count. All 18 solves satisfy the recomputed FP64 residual criterion; no configuration produces a missed escalation.}
\label{tab:replication}
\small
\setlength{\tabcolsep}{6pt}
\begin{tabular}{@{}llccc@{}}
\toprule
State & \(q\) & Reference, FP64 & Replication, FP64 & Replication, FP32 fine \\
\midrule
Cantilever, seed 43 & 0.10 & \(10^{-3}\), 126 & \(10^{-3}\), 126 & \(10^{-3}\), 126 \\
Cantilever, seed 43 & 0.20 & \(10^{-3}\), \phantom{0}69 & \(10^{-3}\), \phantom{0}69 & \(10^{-3}\), \phantom{0}69 \\
Cantilever, seed 43 & 0.35 & \(10^{-12}\), \phantom{0}30 & \(10^{-12}\), \phantom{0}30 & \(10^{-12}\), \phantom{0}30 \\
Bridge, seed 43 & 0.10 & \(10^{-2}\), 196 & \(10^{-2}\), 196 & \(10^{-2}\), 196 \\
Bridge, seed 43 & 0.20 & \(10^{-2}\), 180 & \(10^{-2}\), 179 & \(10^{-2}\), 180 \\
Bridge, seed 43 & 0.35 & \(10^{-3}\), 150 & \(10^{-3}\), 150 & \(10^{-3}\), 149 \\
\bottomrule
\end{tabular}
\end{table}

\section{Discussion}\label{sec:discussion}

\subsection{What the evidence supports}

The evidence supports two separable layers. For this FGMRES implementation at \(\tau=10^{-6}\), a projected-residual stopping flag is not sufficient for acceptance; recomputing the equilibrium residual prevents false acceptance at the cost of one operator application. The same acceptance question is relevant to any Krylov workflow that stops on an inexpensive residual estimate.

The residual features serve a different role: they triage which floor to try first, while the guard retains authority over acceptance. Preserving an admissible original floor avoids mean gradient changes of \(0.340\) and \(0.555\) on severe random states at fixed floors \(10^{-3}\) and \(10^{-2}\), respectively, but only \(0.008\) and \(0.031\) on seven optimized states. The ladder is therefore a state- and model-fidelity choice, not the correctness condition.

\subsection{When is the guarded policy worth its cost?}

The policy is slower than the most plausible fixed-floor alternative. On the held-out test set, a fixed \(10^{-3}\) floor averages 16.2~s per state, versus 39.8~s for the guarded policy. In the single in-loop cantilever comparison, the guarded and fixed-\(10^{-3}\) trajectories end at compliances 2.9620 and 2.9618, while costing 227.3~s and 13.6~s, respectively. That example shows a large certification cost without evidence of a different final design.

For solver implementations that stop on projected residual estimates, acceptance should be based on a freshly recomputed equilibrium residual. The preservation-first ladder is justified when changing the floor changes reported compliance, design sensitivities, or floor-sensitive physics such as buckling \citep{dalklint2021buckling,zhang2021nonersatz}. If the floor is solely a numerical regularization and its modeling effect is acceptable, a fixed \(10^{-3}\) floor with the same residual guard is much cheaper on the tested cases. The present in-loop policy re-probes and re-solves at every design iteration; reusing the previous decision until the guard rejects it is a promising cost reduction that remains untested.

\subsection{Mechanism}\label{sec:mechanism}

The evidence supports an association between coefficient contrast and hierarchy effectiveness, not a unique mechanism. Sparse random fields exhibit a floor-dependent conditioning boundary in reduced direct solves (Figure~\ref{fig:si-atlas}); in the full solver, high early residuals and stalled plateaus provide useful short-run features. The ladder and ablation show that the observed admissibility boundary depends on both the floor and the fine-level adaptive correction. Connectivity alone is insufficient, because the optimizer-generated bridge field is connected and still inadmissible. Among the eight optimizer-generated fields, the only two carrying intermediate densities are the only two hard states, and the ordering is consistent across the set; with eight states this is an ordering observation, not a fitted relationship, and the source-run quality of the two gray fields is itself uneven, so it is not used as a predictive claim.

The reduced direct diagnostic of Section~\ref{sec:conditioning} isolates the floor--conditioning relation but not the floor--multigrid mechanism. The full-solver boundary reflects additional hierarchy- and state-dependent effects, including the fine-level correction identified by the ablation.

The false acceptances of Section~\ref{sec:false-keeps} establish a finite-precision divergence between the projected stopping estimate and a freshly recomputed residual. In exact arithmetic the projected estimate follows the Arnoldi relation; in finite precision it can drift from the explicit residual \citep{saad1993fgmres,carson2017new,higham2022mixed}. High coefficient contrast and the variable preconditioner are plausible contributors, and the fixed-preconditioner control of Section~\ref{sec:control} points to the latter: with the adaptive components disabled, none of the 18 floor attempts produced a false acceptance, and every failure was visible. The FGMRES recurrence itself was not instrumented, so the data localize the effect to the iterate-dependent preconditioner without identifying loss of orthogonality, residual-gap accumulation, or preconditioner variability as the specific path.

\subsection{Limitations}\label{sec:limits}

\emph{Precision was varied on one small set only.} Section~\ref{sec:replication} shows that an FP32 fine-level smoother reproduces the FP64 decisions and iteration counts on six states, but the full held-out test set, other reduced-precision levels, and BF16 were not tested. Whether more aggressive reduced precision changes the frequency of missed escalations is unknown.

\emph{The perturbation evidence is frozen-state.} Both perturbation studies compare compliance and gradients at a fixed density field. They do not measure what a raised floor does to an optimization \emph{trajectory}, where small per-iteration gradient changes could accumulate or cancel; the fixed-floor trajectory controls of Section~\ref{sec:in-loop} hint at the question but carry no acceptance record and are three runs per geometry.

\emph{Scope.} The evidence covers one solver implementation, one discretization family, structured hexahedral 3D elasticity, and 40,500 to 514,500 elements on one reference GPU, plus one \(10^6\)-element state in the replication configuration. The monotone floor argument proves no convergence result for an arbitrary hierarchy, and the reduced direct-solve conditioning experiment does not identify the multigrid mechanism. Thresholds, ladder values, and timings are implementation measurements. Unstructured meshes, nonlinear or nearly incompressible response, contact-like boundary conditions, and larger meshes may change them. Two in-loop trajectories at one resolution support no claim about optimization paths or final topologies.

\section{Conclusion}

For the tested matrix-free GMG-FGMRES implementation, a positive SIMP floor does not ensure that the projected-residual stopping test and the recomputed equilibrium residual agree. The mismatch occurs on four of 102 held-out states and propagates silently through 22 of 40 outer iterations in one unguarded trajectory. A 100-iteration probe correctly triages 98 of 102 reference classifications, but the recomputed-residual guard is the correctness component: its acceptance test costs one additional operator application and catches all four missed escalations; rejected states then incur the reported fallback-solve cost. The floor ladder then controls a separate fidelity--cost tradeoff. The reduced-operator measurements additionally yield a hierarchy-independent screening rule, \(\rhoMin\gtrsim c(\rr)\varepsilon/\tau\), below which no backward-stable FP64 solve can be guaranteed to certify the tolerance, conservative on all 18 tested fields. Relative to a fixed \(10^{-3}\) floor, preserving the original operator avoids mean compliance and gradient changes of \(31.0\%\) and \(0.340\) on 24 severe random states, and \(0.48\%\) and \(0.008\) on seven optimized states, while increasing mean wall time by a factor of \(2.5\). The general recommendation is therefore scoped: recompute the equilibrium residual whenever acceptance otherwise rests on a projected estimate; use adaptive floor preservation when the modeling effect of a fixed raised floor justifies its additional cost. The numerical thresholds and ladder values remain specific to the tested solver and state families.

\section*{Data Availability}

The code implementing the solver stack, the guarded policy, and every experiment reported here is available under a BSD 3-Clause licence at\\
\url{https://github.com/nbbllxx0/solver-admissibility-and-guarded-floor-selection}. The repository contains the matrix-free operator, the multigrid hierarchy, all experiment drivers and analyzers, the figure generators, and pinned environment files for both platforms in Table~\ref{tab:platforms}, together with the exact commands that reproduce each figure and table.

The research-data package contains the analysis tables and the eight optimized-density inputs required for the transfer study and Figure~\ref{fig:si-gallery}. It will be deposited in a public research-data repository before journal submission, and the persistent identifier will be added here. All results except the transfer study and hardware-specific wall-time tables can be regenerated from the code repository alone.

\section*{Acknowledgment}

This work was supported by Santa Clara University.

\section*{Declaration of Competing Interest}

The authors declare that they have no known competing financial interests or personal relationships that could have appeared to influence the work reported in this paper.

\section*{Declaration of generative AI and AI-assisted technologies in the manuscript preparation process}

During preparation of this manuscript, the authors used Claude (Anthropic) and OpenAI Codex to assist with language, clarity, and manuscript organization. The authors reviewed and revised all AI-assisted material and retain full responsibility for the manuscript's content, analyses, and conclusions.

\section*{CRediT Author Statement}

Shaoliang Yang: Conceptualization, Methodology, Software, Validation, Formal analysis, Investigation, Visualization, Writing (original draft). Jun Wang: Conceptualization, Methodology, Supervision, Writing (review and editing). Yunsheng Wang: Supervision, Methodology, Writing (review and editing).

\appendix

\section{Reproducibility}\label{app:repro}
\setcounter{table}{0}
\renewcommand{\thetable}{A.\arabic{table}}
\renewcommand{\theHtable}{A.\arabic{table}}

\paragraph{Environment.}
Two computational configurations were used (Table~\ref{tab:platforms}). Every reported timing and all primary evidence come from the reference configuration, pinned in \texttt{environment.yml}: Python~3.10.18, CuPy~13.6.0, NumPy~2.2.6, SciPy~1.15.3, Matplotlib~3.10.7, pandas~2.3.3, scikit-image~0.25.2, PyVista~0.46.3, pyamg~5.3.0, on one NVIDIA GeForce RTX~4090. The replication configuration differs in GPU architecture, CuPy major version, and CUDA major version; it is used for the replication and precision checks of Section~\ref{sec:replication}, the optimized-design perturbation study of Section~\ref{sec:perturb}, and the million-element scale check of Section~\ref{sec:transfer}. Wall times are not compared across configurations.

\begin{table}[H]
\centering
\caption{Computational configurations. Acceptance decisions and iteration counts agree within the differences reported in Section~\ref{sec:replication}; wall times are not compared across configurations.}
\label{tab:platforms}
\small
\setlength{\tabcolsep}{5pt}
\begin{tabularx}{0.98\linewidth}{@{}p{0.16\linewidth}p{0.26\linewidth}p{0.24\linewidth}Y@{}}
\toprule
Configuration & GPU & Software & Used for \\
\midrule
Reference & GeForce RTX 4090, 24~GiB, SM~8.9 & CuPy 13.6.0, CUDA 12.x, Python 3.10 & Primary evidence and all timing comparisons. \\
Replication & GeForce RTX 5090, 32~GiB, SM~12.0 & CuPy 14.0.1, CUDA 13.0, Python 3.11 & Six-state solver-stack replication, FP32 fine-smoother check, optimized-design perturbation study, and million-element scale check. \\
\bottomrule
\end{tabularx}
\end{table}

\paragraph{Cost and scope of the study.}
The fully timed 102-state policy comparison accounts for 6.47 GPU-hours; additional partially timed diagnostics make the total larger. Energy and emissions were not benchmarked. The study uses synthetic random density fields and stored mechanical optimization states only; it contains no human or personal data.

\paragraph{Reference classifications and selected-floor records.}
The binary reference classification and the selected-floor record answer different questions. Six bridge states (seeds 43 and 47 at \(q=0.10,0.15,0.20\)) are classified as requiring escalation and receive \(10^{-2}\) after the reference \(10^{-12}\) solve exhausts its budget, yet converge from a fresh \(10^{-3}\) attempt in the guarded path. Reference-classification counts and selected-floor counts are therefore not the same statistic.

\paragraph{Krylov basis memory at \(10^6\) elements.}
Flexible GMRES stores two Krylov bases, so at the \(3.1\times10^6\) free degrees of freedom of the \(10^6\)-element cantilever state of Section~\ref{sec:transfer} each outer iteration adds about \(50\)~MiB of FP64 basis. The accepted 50-iteration solve therefore peaks at \(32.3\)~GiB of the \(32.6\)~GiB available on the replication GPU, and a state consuming the full \(k_{\max}=300\) budget would need roughly \(12\)~GiB more than the card provides. A restarted variant (restart 100, \(5.0\)~GiB of basis) is included in the code release for full-budget runs at that size.

\section{Supplementary Material}\label{app:supp}
\setcounter{table}{0}
\renewcommand{\thetable}{S\arabic{table}}
\renewcommand{\theHtable}{S.\arabic{table}}
\setcounter{figure}{0}
\renewcommand{\thefigure}{S\arabic{figure}}
\renewcommand{\theHfigure}{S.\arabic{figure}}

\subsection{Case taxonomy and trajectory settings}

\begin{table}[H]
\centering
\caption{Case families used in the evaluation.}
\label{tab:case-taxonomy}
\footnotesize
\setlength{\tabcolsep}{3pt}
\begin{tabularx}{0.98\linewidth}{L{0.19\linewidth}L{0.22\linewidth}L{0.30\linewidth}Y}
\toprule
Family & Setup & Density source & Role \\
\midrule
Reduced conditioning diagnostic & Reduced cantilever, \(24\times12\times6\) & Bernoulli fields, seeds 7/13/19, \(q=0.10\)--0.35, generated with NumPy \texttt{default\_rng(seed)} and threshold \(u<q\) & Preliminary evidence that a floor boundary exists before the full hierarchy is used \\
Development states & Cantilever, 64k & 16 difficult fields identified before testing, seeds 7, 13 and 19 with \(q=0.10\)--0.35 & Threshold selection only \\
Held-out test states & Cantilever and bridge, 40.5k--64k & 102 states, cantilever seeds 41--71 with \(q=0.08\)--0.35, bridge seeds 41--59 with \(q=0.10\)--0.35 & Primary evaluation; not used for threshold selection \\
Stage-cost states & Cantilever and bridge, 40.5k--64k, plus four optimized fields & 19 states: cantilever seeds 23 and 31, bridge seed 23, and the 64k/512k cantilever, 512k bracket and 514.5k bridge designs & Per-stage cost accounting (Section~\ref{sec:cost}); disjoint from the random development and test sets \\
Optimizer-generated states & Six geometries, 64k--514.5k & Eight unique stored final SIMP density fields & Transfer beyond synthetic random states \\
In-loop trajectories & Cantilever 64k, bridge 40.5k & Live 40-iteration optimization runs & Consequences of the acceptance guard during optimization \\
\bottomrule
\end{tabularx}
\end{table}

\begin{table}[H]
\centering
\caption{Optimizer settings for the 40-iteration trajectories of Section~\ref{sec:in-loop}. These apply only to the trajectory runs; frozen-state experiments use \(p_{\mathrm{SIMP}}=4.5\).}
\label{tab:trajectory-settings}
\small
\setlength{\tabcolsep}{3pt}
\begin{tabularx}{0.95\linewidth}{p{0.26\linewidth}Y}
\toprule
Setting & Value \\
\midrule
Formulation & The density-based compliance problem of Section~\ref{sec:formulation} with the optimality-criteria update, run under the continuation schedule below; every equilibrium solve uses the stack of Table~\ref{tab:solver-config}. \\
Continuation & Outer iterations 1--15: \(p_{\mathrm{SIMP}}=1.5\), \(\beta=1\), move limit 0.20. Iterations 16--40: \(p_{\mathrm{SIMP}}=3.5\), \(\beta=4\), move limit 0.15. \\
Filter and volume & Filter radius \(r_{\min}=1.5\) elements; volume fraction 0.30 for both geometries. \\
Budget & 40 outer iterations, minimum 20, optimization tolerance 0.01. \\
Floor variants & Fixed \(\rhoMin\in\{10^{-12},10^{-3},10^{-2}\}\) without acceptance testing; guarded runs select among \(\rho_0\) and the ladder using Algorithm~\ref{alg:floor_policy} at every outer iteration. \\
\bottomrule
\end{tabularx}
\end{table}

\begin{table}[H]
\centering
\caption{Source states for the optimized-density transfer of Section~\ref{sec:transfer}; domains, meshes, supports and loads for each family are in Table~\ref{tab:geometries}. \(C_{\mathrm{src}}\) and the grayness are the values recorded by the source run. ``At cap'' is the percentage of that run's 120 outer iterations whose state solve terminated at the source solver's iteration cap rather than at a tolerance. For the two fields at \(100\%\) the recorded compliance also disagrees with the last logged iterate --- for the MBB beam, 2.871 recorded against 3.124 logged --- so those stored states were selected on compliance values computed from solves that never met a tolerance.}
\label{tab:optimized-source-metadata}
\scriptsize
\setlength{\tabcolsep}{2pt}
\begin{tabularx}{0.98\linewidth}{L{0.20\linewidth}r r r r Y}
\toprule
Family & Vol. & \(C_{\mathrm{src}}\) & Grayness & At cap & Source-state settings \\
\midrule
64k cantilever & 0.300 & 2.865 & \(1.03{\times}10^{-8}\) & 1\% & \(p=4.5\), \(r_{\min}=1.2\), \(\beta=32\), move 0.05 \\
216k cantilever & 0.300 & 2.172 & \(9.27{\times}10^{-6}\) & 1\% & \(p=4.5\), \(r_{\min}=1.2\), \(\beta=32\), move 0.05 \\
512k cantilever & 0.300 & 1.825 & \(3.08{\times}10^{-5}\) & 3\% & \(p=4.5\), \(r_{\min}=1.2\), \(\beta=32\), move 0.05 \\
512k bracket & 0.300 & 0.820 & \(1.58{\times}10^{-6}\) & 4\% & \(p=4.5\), \(r_{\min}=1.2\), \(\beta=32\), move 0.05 \\
499k torsion & 0.250 & 2.036 & \(1.00{\times}10^{-5}\) & 92\% & \(p=4.5\), \(r_{\min}=1.2\), \(\beta=32\), move 0.05 \\
500k column & 0.200 & 0.249 & \(3.99{\times}10^{-5}\) & 84\% & \(p=4.5\), \(r_{\min}=1.2\), \(\beta=32\), move 0.05 \\
514.5k MBB beam & 0.500 & 2.871 & \(6.60{\times}10^{-3}\) & \textbf{100\%} & \(p=4.5\), \(r_{\min}=1.25\), \(\beta=16\), move 0.08 \\
514.5k bridge & 0.300 & 1.236 & \(9.43{\times}10^{-4}\) & \textbf{100\%} & \(p=4.5\), \(r_{\min}=1.25\), \(\beta=16\), move 0.08 \\
\bottomrule
\end{tabularx}
\end{table}

\begin{table}[H]
\centering
\caption{Fixed-preconditioner control of Section~\ref{sec:control}: all nine states and all 18 floor attempts, run in the replication configuration with both adaptive preconditioner components disabled and the escalation rule switched off, so the original floor is always attempted to exhaustion. ``Projected \(\eta\)'' is the stopping estimate at termination of a rejected attempt; the two bridge attempts that terminate before the 300-iteration cap exit by solver breakdown, with estimates near unity. No attempt stopped with its projected estimate below the tolerance while the recomputed residual failed, so no false acceptance occurred.}
\label{tab:control}
\scriptsize
\setlength{\tabcolsep}{3.5pt}
\begin{tabularx}{0.98\linewidth}{@{}llYlrr@{}}
\toprule
State & \(q\) & Rejected attempts: floor (iterations, projected \(\eta\)) & Accepted floor & Iterations & True residual \\
\midrule
Cantilever, seed 47 & 0.30 & \(10^{-12}\) (300, \(5.3{\times}10^{-6}\)) & \(10^{-3}\) & 42 & \(8.4{\times}10^{-7}\) \\
Cantilever, seed 47 & 0.35 & \(10^{-12}\) (300, \(2.8{\times}10^{-6}\)) & \(10^{-3}\) & 15 & \(4.6{\times}10^{-7}\) \\
Cantilever, seed 59 & 0.30 & \(10^{-12}\) (300, \(8.0{\times}10^{-4}\)) & \(10^{-3}\) & 36 & \(5.0{\times}10^{-7}\) \\
Cantilever, seed 59 & 0.35 & none & \(10^{-12}\) & 26 & \(6.4{\times}10^{-7}\) \\
Cantilever, seed 71 & 0.30 & \(10^{-12}\) (300, \(2.8{\times}10^{-4}\)) & \(10^{-3}\) & 41 & \(8.2{\times}10^{-7}\) \\
Cantilever, seed 71 & 0.35 & \(10^{-12}\) (300, \(4.0{\times}10^{-6}\)) & \(10^{-3}\) & 15 & \(7.3{\times}10^{-7}\) \\
Cantilever, seed 43 & 0.35 & none & \(10^{-12}\) & 130 & \(9.7{\times}10^{-7}\) \\
Bridge, seed 43 & 0.10 & \(10^{-12}\) (208, \(1.0{\times}10^{0}\)); \(10^{-3}\) (300, \(9.8{\times}10^{-1}\)); \(10^{-2}\) (300, \(9.9{\times}10^{-1}\)) & none & --- & --- \\
Bridge, seed 43 & 0.20 & \(10^{-12}\) (112, \(9.7{\times}10^{-1}\)); \(10^{-3}\) (300, \(9.6{\times}10^{-1}\)); \(10^{-2}\) (300, \(9.9{\times}10^{-1}\)) & none & --- & --- \\
\bottomrule
\end{tabularx}
\end{table}

\subsection{Supplementary figures}

\begin{figure}[H]
  \centering
  \includegraphics[width=0.96\linewidth]{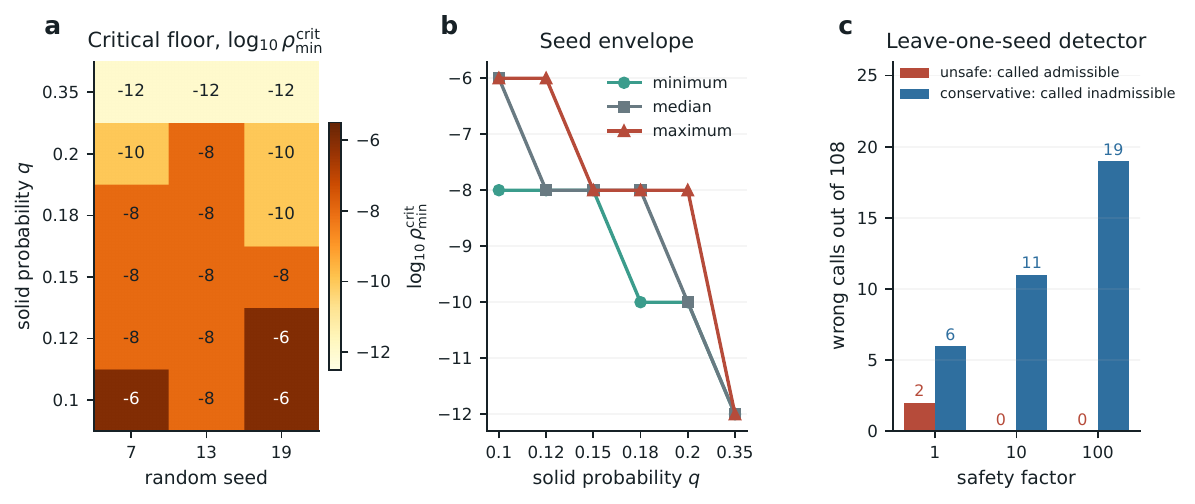}
  \caption{Reduced direct diagnostic on \(24\times12\times6\) cantilever fields, used only to establish that a floor boundary exists before the full hierarchy is considered. (a) Critical floor per seed and solid probability, labelled with \(\log_{10}(\rho_{\min}^{\mathrm{crit}})\). (b) Seed-wise envelope. (c) Leave-one-seed detector errors against the safety factor: at factor 10 there are no unsafe false-admissible calls out of 108 predictions, at the cost of 11 conservative ones.}
  \label{fig:si-atlas}
\end{figure}

\begin{figure}[H]
  \centering
  \includegraphics[width=0.92\linewidth]{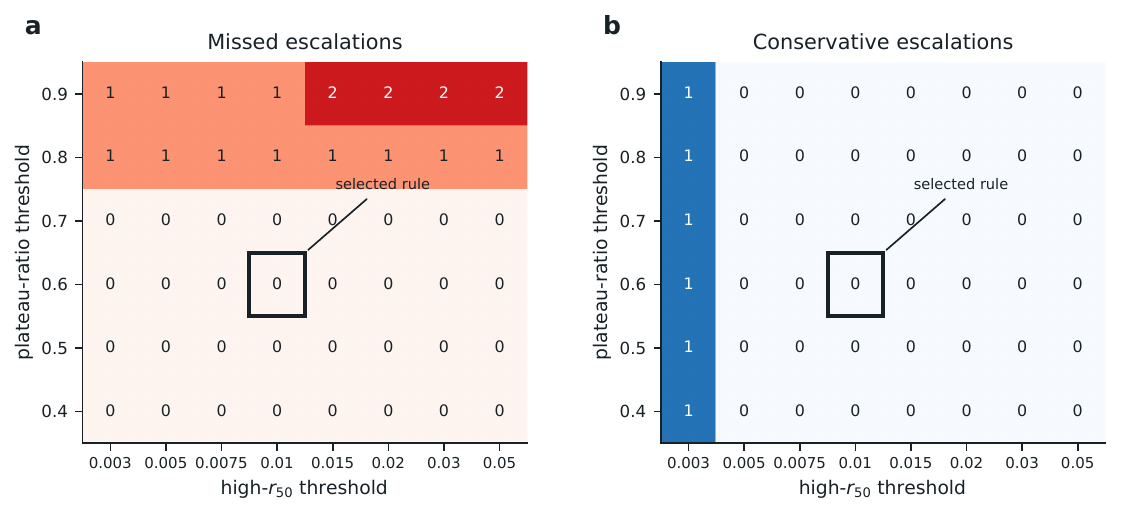}
  \caption{Threshold sensitivity of the preserve/escalate rule at plateau-residual threshold \(10^{-4}\), one slice of the 288-combination sweep on the 16 development states. (a) Missed escalations; (b) conservative escalations. Cell values are state counts. The square marks the reported rule, which has no missed escalation in this displayed slice. Results aggregated over all six plateau-residual slices are reported in Section~\ref{sec:robust}.}
  \label{fig:si-threshold}
\end{figure}

\begin{figure}[H]
  \centering
  \includegraphics[width=0.98\linewidth]{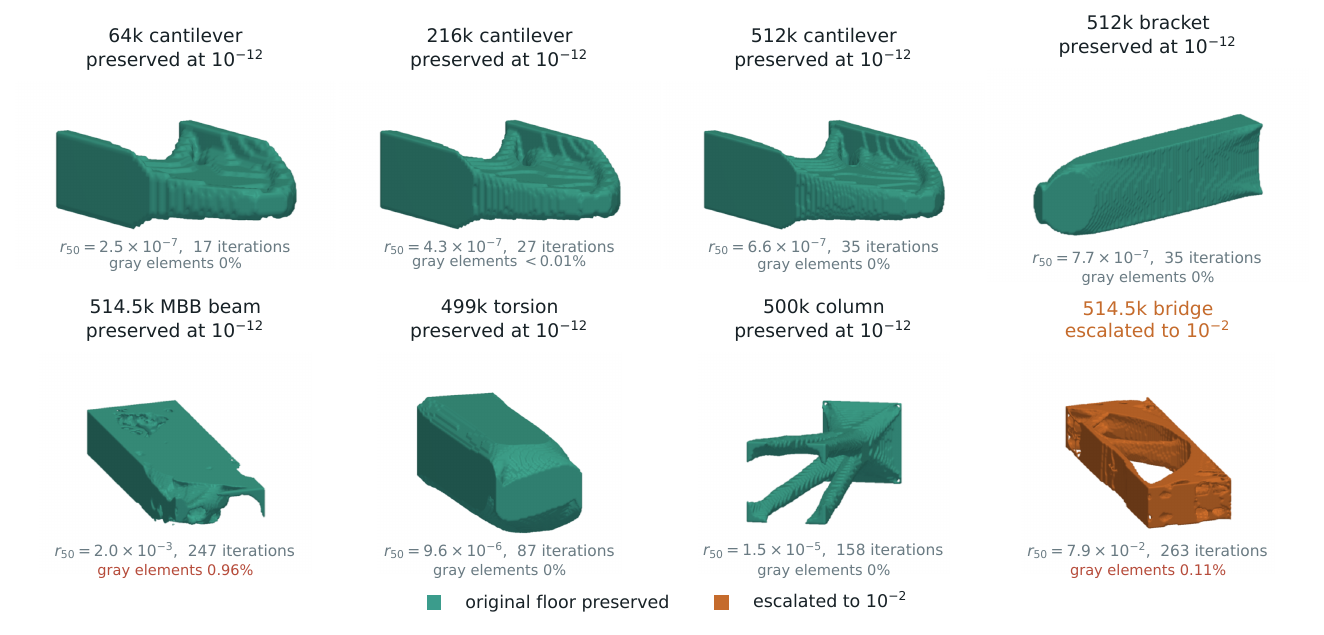}
  \caption{The eight unique optimized density fields of Table~\ref{tab:optimized-transfer}, rendered as orthographic isosurfaces at \(\rho\ge0.5\) and viewed from the front upper right of each domain as oriented in Table~\ref{tab:geometries}. Colour gives the policy decision: teal preserves the original floor, orange requires escalation. Each panel reports the probe feature \(r_{50}\), the accepted-solve iteration count, and the fraction of elements with \(0.05<\widehat{\rho}_e<0.95\) (red where that fraction exceeds \(0.05\%\)). Fields are shown exactly as stored, including the two whose source runs terminated every state solve at the iteration cap; Section~\ref{sec:transfer} discusses what that means for the MBB beam, whose isosurface is a diffuse slab rather than a resolved truss.}
  \label{fig:si-gallery}
\end{figure}

\begin{figure}[H]
  \centering
  \includegraphics[width=0.92\linewidth]{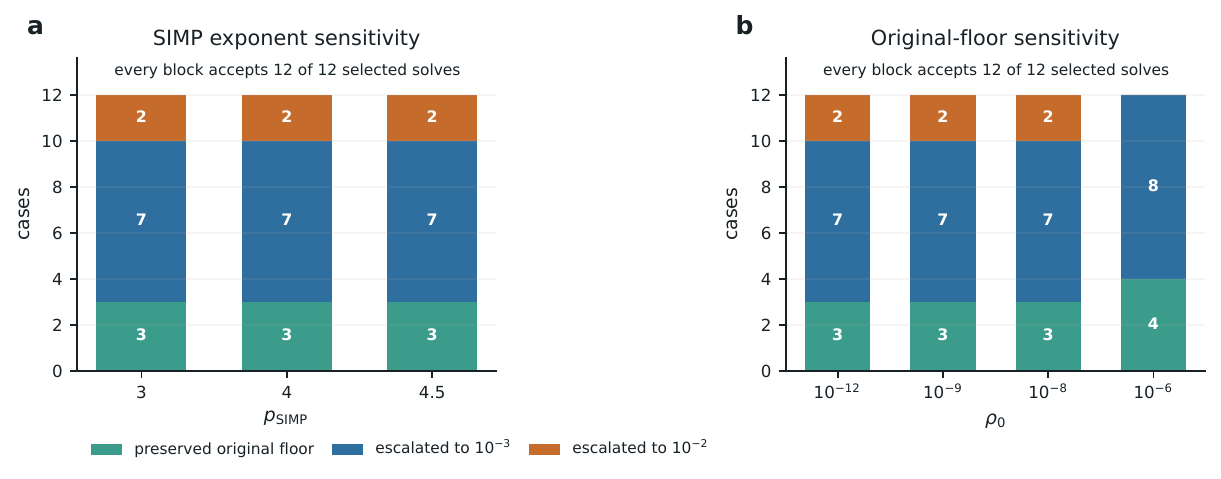}
  \caption{Sensitivity of the guarded policy on the 12 states defined in Section~\ref{sec:robust}; segment labels are state counts. (a) SIMP exponent \(p_{\mathrm{SIMP}}\in\{3,4,4.5\}\) gives identical selected-floor counts. (b) Original floor \(\rho_0\in\{10^{-12},10^{-9},10^{-8}\}\) reproduces the same pattern, while \(\rho_0=10^{-6}\) preserves the original floor on four states and escalates eight to \(10^{-3}\). Every selected solve satisfies the recomputed-residual test.}
  \label{fig:si-sensitivity}
\end{figure}

\FloatBarrier

\bibliographystyle{unsrtnat}
\bibliography{refs/references}

\end{document}